\documentclass[a4paper,aps,showpacs,groupedaddress,pre,twocolumn]{revtex4-1}

\usepackage{amsmath}
\usepackage{graphicx}
\usepackage{color}

\newcommand{\be}{\begin{equation}}
\newcommand{\ee}{\end{equation}}
\newcommand{\bea}{\begin{eqnarray}}
\newcommand{\eea}{\end{eqnarray}}

\newcommand{\cs}{c_{\text{s}}}
\newcommand{\Ns}{N_{\text{s}}}

\begin{document}

\title{Semi--automatic construction of Lattice Boltzmann models}

\author{Dominic Spiller}

\affiliation{Max Planck Institute for Polymer Research, Ackermannweg
  10, 55128 Mainz, Germany}

\author{Burkhard D\"unweg}

\affiliation{Max Planck Institute for Polymer Research, Ackermannweg
  10, 55128 Mainz, Germany}

\affiliation{Department of Chemical Engineering, Monash University,
  Clayton, Victoria 3800, Australia}

\date{\today}

\begin{abstract}
  A crucial step in constructing a Lattice Boltzmann model is the
  definition of a suitable set of lattice velocities, and the correct
  assignment of the associated weights. For high--order models, the
  solution of this problem requires a non--trivial effort. The paper
  outlines the functioning of a publicly available Python script which
  has been written to assist researchers in that task. The speed of
  sound $\cs$ is considered as a parameter, which can, within limits,
  be chosen at will. Under this premise, the Maxwell--Boltzmann
  constraint equations are a system of linear equations to determine
  the weights, and hence amenable to numerical solution by standard
  linear algebra library routines. By suitable contractions, the
  tensor equations are mapped to a set of equivalent scalar equations,
  which simplifies the treatment significantly. For a user--supplied
  set of velocity shells, the software first checks if a solution
  for the weights exists, and returns it if it also happens to be
  unique. In such a case, the software also calculates the range of
  $\cs$ values that yield positive weights. Standard models like D3Q19
  with a well--defined special $\cs$ value then result as limiting
  cases where one of the weights vanishes. In case of an infinite set
  of solutions, the user may find one particular solution by supplying
  a $\cs$ value, and then minimizing one or several weights within the
  framework of standard linear programming. Some examples illustrate
  the feasibility and usefulness of the approach. A number of models
  that have been discussed in the literature are nicely reproduced,
  while the software has also been able to find some new models of
  even higher order.
\end{abstract}


\maketitle

\section{Introduction: General background, and
  definition of the problem}
\label{sec:intro}

The Lattice Boltzmann (LB) method~\cite{succi_lattice_boltzmann_2001,
  succi_lattice_2018, kruger_lattice_2016, benzi_lattice_1992,
  dunweg_lattice_2009} can nowadays be viewed as a mature and
well--established method to solve the equations of motion of fluid
dynamics. Briefly, the method is based upon a regular lattice, each of
whose sites $\vec r$ at time $t$ contains a finite set of populations
$n_i (\vec r, t)$. The index $i$ is associated with a corresponding
finite set of velocities (or lattice speeds) $\vec c_i$. This set is
chosen commensurate with the symmetry of the lattice. The velocities
are used for the streaming step of the algorithm, where $n_i(\vec r,
t)$ is, within one time step $h$, moved to a new site $\vec r' = \vec
r + h \vec c_i$:
\be
  n_i(\vec r + h \vec c_i, t + h) = n_i(\vec r, t) .
\ee
In other words, the velocities must be chosen in such a way that they
carry the populations from one site to another (and not to some
``interstitial site''). Interactions are modeled by an additional
collision step, where $\Delta_i (\vec r, t)$ is the so--called
``collision operator'', such that the full update rule (the
so--called Lattice Boltzmann Equation (LBE)) is given by
\be
  n_i(\vec r + h \vec c_i, t + h) = n_i(\vec r, t)
  + \Delta_i (\vec r, t) .
\ee
The populations are usually identified with the mass densities
associated with their corresponding velocities, such that the
total mass density $\rho$ at the local site is given by
\be
  \rho (\vec r, t) = \sum_i n_i (\vec r, t) .
\ee
Similarly, the momentum density $\vec j$ is given by
\be
  \vec j (\vec r, t)  = \sum_i n_i (\vec r, t) \vec c_i =
  \rho (\vec r, t) \vec u (\vec r, t) ,
\ee
where $\vec u (\vec r, t)$ is the local streaming velocity. The
collision operator is then constructed in such a way that it locally
conserves the mass density,
\be
  \sum_i \Delta_ i = 0 ,
\ee
as well as the momentum density,
\be
  \sum_i \Delta_i \vec c_i = 0.
\ee
An additional conservation law for the kinetic energy may be added if
the method is intended to not only simulate isothermal hydrodynamics,
but also heat transport.

In what follows, we will assume that the lattice is a simple cubic
lattice in $d$ spacial dimensions. We will also use natural units,
where both the lattice spacing as well as the time step $h$ are set to
unity.

The standard and most popular version of the LBE is based upon a
\emph{linearized} Boltzmann equation~\cite{higuera_lattice_1989,
  qian_lattice_1992}. In terms of a ``cookbook recipe'' it may be
described as follows: One first obtains the local conserved quantities
$\rho$ and $\vec j$ (and $\vec u = \vec j / \rho$), which are then
used to calculate a set of local equilibrium populations:
\be
  \label{eq:equilibrium}
  n_i^{eq} (\rho, \vec u) = w_i \rho 
  \left( 1 + \frac{\vec u \cdot \vec c_i}{\cs^2} 
  + \frac{(\vec u \cdot \vec c_i)^2}{2 \cs^4} -
  \frac{u^2}{2 \cs^2} \right) .
\ee
Here $\cs$ denotes the (isothermal) speed of sound, while the
coefficients $w_i$ are a set of positive weights associated with the
velocities $\vec c_i$. For symmetry reasons, these weights must take
the same value within a velocity shell. Here a shell is defined as the
equivalence class of all lattice speeds that can be mapped onto each
other by one of the symmetry operations of the lattice's point group,
see also Sec.~\ref{sec:VelocityShells}. Furthermore, we require the
moment conditions
\begin{eqnarray}
  \label{eq:tensor0}
  && \sum_i w_i = 1 , \\
  \label{eq:tensor1}
  && \sum_i w_i \vec c_i = 0 , \\
  \label{eq:tensor2}
  && \sum_i w_i c_{i\alpha} c_{i \beta}
  = \cs^2 \delta_{\alpha \beta} , \\
  \label{eq:tensor3}
  && \sum_i w_i c_{i \alpha} c_{i \beta} c_{i \gamma}
  = 0 , \\
  \label{eq:tensor4}
  \nonumber
  && \sum_i w_i c_{i \alpha} c_{i \beta} c_{i \gamma}
  c_{i \delta} = \\
  && \cs^4 \left( \delta_{\alpha \beta}
  \delta_{\gamma \delta} + \delta_{\alpha \gamma}
  \delta_{\beta \delta} + \delta_{\alpha \delta}
  \delta_{\beta \gamma} \right)  ,
\end{eqnarray}
where Greek letters denote Cartesian indexes, for which the Einstein
summation convention is implied. It should be noted that
Eqs.~\ref{eq:tensor1} and \ref{eq:tensor3} are valid automatically for
symmetry reasons. Similarly, the only aspect of Eq.~\ref{eq:tensor2}
that does not follow automatically from symmetry is the value of the
prefactor of the unit tensor on the right hand side (rhs). In
contrast, Eq.~\ref{eq:tensor4} is less trivial: Not only is there a
need to adjust the prefactor $\cs^4$ on the rhs, but we also need to
ensure that the fourth--rank tensor is isotropic: From cubic symmetry
alone, the form of the rhs is not guaranteed at all --- rather one
expects an additional term $\kappa_4 \delta_{\alpha \beta \gamma
  \delta}$, where $\delta_{\alpha \beta \gamma \delta}$ is one for all
indexes being the same, and zero otherwise. Therefore one needs to
adjust the coefficients in such a way that $\kappa_4$ vanishes. The
well--known D3Q19 model~\cite{qian_lattice_1992} is one possible
solution of this problem: Here the velocities on the
three--dimensional cubic lattice comprise the three shells with
$\vec{c_i}^2 = 0, 1, 2$ (one velocity $+$ six velocities $+$ twelve
velocities $= 19$ velocities), and the weights are given by $w_i =
1/3, 1/18, 1/36$, respectively, for the three shells. For this model,
the speed of sound takes the value $\cs^2 = 1/3$.

Via straightforward calculation one then shows that the equilibrium
populations according to Eq.~\ref{eq:equilibrium} satisfy analogous
moment conditions:
\begin{eqnarray}
  \label{eq:equiltensor0}
  \sum_i n_i^{eq} & = & \rho , \\
  \label{eq:equiltensor1}
  \sum_i n_i^{eq} \vec c_i & = & \vec j , \\
  \label{eq:equiltensor2}
  \sum_i n_i^{eq} c_{i \alpha} c_{i \beta} & = &
  \rho \cs^2 \delta_{\alpha \beta} + \rho u_\alpha u_\beta .
\end{eqnarray}
It should be noted that the model implies the thermodynamics of an
ideal gas. If $m$ denotes the mass of a gas particle, the equation of
state is given by
\be
  p = \frac{\rho}{m} k_B T ,
\ee
where $p$ is the thermodynamic pressure, $k_B$ Boltzmann's constant,
and $T$ the absolute temperature. Since the speed of sound is given by
$\cs^2 = \partial p / \partial \rho$, it is clear that $\rho \cs^2$ in
Eq.~\ref{eq:equiltensor2} is indeed just the pressure, such that the
whole rhs of Eq.~\ref{eq:equiltensor2} is just the Euler stress
occurring in the Navier--Stokes equation. Furthermore, we note that,
for an ideal gas which is globally at rest, the kinetic energy of a
gas particle, in units of $k_B T$, can be written as
\be
  \frac{\frac{m}{2} \vec v^2}{k_B T} =
  \frac{\vec v^2}{2} \frac{\rho}{p} =
  \frac{\vec v^2}{2 \cs^2} ,
\ee
where $\vec v$ is the particle velocity.

After obtaining the equilibrium populations as discussed, one then
constructs a linearized collision operator
\be
\Delta_i = - \sum_j L_{ij} \left( n_j - n_j^{eq} \right) ,
\ee
where the coefficients $L_{ij}$ encode details about the dissipative
processes in the system (i.~e. viscous damping in isothermal
hydrodynamics). Via a Chapman--Enskog expansion (see, e.~g.,
Ref.~\cite{dunweg_lattice_2009} for details) one then shows that for
small Mach numbers (i.~e. ignoring terms of order $(u/\cs)^3$)
Navier--Stokes dynamics is recovered in the continuum limit.

Considering the continuum statistical mechanics of the gas at rest
($\vec u = 0$), the velocity distribution of the particles is given by
the Maxwell--Boltzmann distribution
\be
  f(\vec v) = \left(2 \pi \cs^2\right)^{-d/2}
  \exp \left( - \frac{\vec v^2}{2 \cs^2} \right) .
\ee
In analogy to Eqs.~\ref{eq:tensor0} to \ref{eq:tensor4} we can
therefore similarly consider the velocity moments
\begin{eqnarray}
  && \int d^d \vec v \, f(\vec v) = 1 , \\
  && \int d^d \vec v \, f(\vec v) \vec v = 0 , \\
  && \int d^d \vec v \, f(\vec v) v_\alpha v_\beta =
  \cs^2 \delta_{\alpha \beta} , \\
  && \int d^d \vec v \, f(\vec v) v_\alpha v_\beta
  v_\gamma = 0 , \\
  \nonumber
  && \int d^d \vec v \, f(\vec v) v_\alpha v_\beta
  v_\gamma v_\delta = \\
  && \cs^4 \left( \delta_{\alpha \beta}
  \delta_{\gamma \delta} + \delta_{\alpha \gamma}
  \delta_{\beta \delta} + \delta_{\alpha \delta}
  \delta_{\beta \gamma} \right)  ,
\end{eqnarray}
which means that we can write the moment conditions
Eqs.~\ref{eq:tensor0}--\ref{eq:tensor4} for the
coefficients $w_i$ in the compact form of so--called
``Maxwell--Boltzmann constraints'' (MBCs)
\be
  \label{eq:MBC}
  \sum_i w_i c_{i \alpha} c_{i \beta} \ldots c_{i \gamma} =
  \int d^d \vec v f(\vec v) v_\alpha v_\beta \ldots
  v_\gamma
\ee
for all tensor ranks up to rank four.

If we ignore the details of the collision operator, and also problems
of stability, accuracy, staggered invariants, etc., we may therefore
say that the construction of a standard LB model is tantamount to the
two steps:
\begin{enumerate}
  \item[LB1:] Find a suitable set of velocities $\vec c_i$; and
  \item[LB2:] calculate the weights $w_i$, based upon satisfying
    Eq.~\ref{eq:MBC}, which is therefore seen to lie at the
    heart of the process.
\end{enumerate}
Of course, it is possible to solve problems LB1 and LB2 merely with
paper--and--pencil work. However, already for the D3Q19 model this is
a task that can no longer be viewed as completely
trivial. Furthermore, we should take into account that there is a
growing trend in the community~\cite{shan_kinetic_2006,
  philippi_continuous_2006, chen_discrete_2008,
  chikatamarla_lattices_2009, karlin_pap07_2010, shan_pap04_2010,
  shan_pap05_2016} to consider higher--order LB models, which means,
in the present context, the study of larger velocity sets with
suitably adjusted weights, such that Eq.~\ref{eq:MBC} is satisfied for
even higher--rank tensors than just fourth order. Except for the goal
to obtain a better degree of isotropy, which is of course desirable as
such, there are also cases where the physics \emph{dictates} such
higher--order models. One example is thermal transport, where the
hydrodynamic equation of motion for the energy density contains a term
$\propto u^3$, such that the expansion of $n_i^{eq}$ in powers of $u$
needs to be carried to higher than second order, which in turn means
that also higher--order velocity moments appear in the
theory~\cite{shan_central-moment-based_2019}.  Even for isothermal
flows, it has been demonstrated that the improved isotropy properties,
which result from a larger velocity set, significantly help in the
removal of artifacts, in particular in problems where rotational
symmetry plays a crucial
role~\cite{white_pap02_2011,silva_pap03_2014}. Yet another example is
the study of isothermal gas--liquid systems within the framework of a
density--functional approach with a smeared--out interface. Here the
interface is modeled by a gradient--square term in the free energy
functional, such that a third--order gradient of density occurs in the
Navier--Stokes equations as an interfacial driving force. Therefore
such a system requires a Chapman--Enskog expansion up to third
order~\cite{wagner_thermodynamic_2006} and, concomitantly, correct
MBCs up to sixth--rank tensors. These issues shall not be further
discussed here. We are rather concerned with the solution of LB1 and
LB2 as such, just as a mathematical problem, which we wish to solve in
a fairly general fashion with maximum use of a computer and minimum
paper--and--pencil work, since the latter is both cumbersome and
error--prone, in particular for high--order models.

It turns out that the problem is most suitable for solution on the
computer if we consider $\cs^2$ not as some ``magic number'' (like
$\cs^2 = 1/3$) resulting from the analysis, but rather as a parameter
that can (within limits) be chosen freely at will. This additional
degree of freedom requires at least one additional velocity shell,
compared to models like D3Q19 with a fixed and prescribed value of
$\cs^2$. At first glance, this might be viewed as an unnecessary
complication; however, the advantage of this treatment is that in this
way the problem becomes strictly \emph{linear}, such that standard
library routines of linear algebra become applicable. Furthermore,
there are cases where the physics of the problem anyway makes it
desirable to have $\cs^2$ available as a free parameter: Since the
equation of state is given by $p = \rho \cs^2$, one can implement a
non--trivial equation of state by making $\cs^2$ a parameter that
depends on the local density. Finally, it should be noted that models
with ``magic'' $\cs^2$ values like D3Q19 can be derived very easily
from the more general treatment: The ``magic'' $\cs^2$ is just the
value that causes the weight of the additional shell to vanish, which
means that this shell simply does not occur in the thus--reduced
model.

The purpose of the present paper is to derive an algorithm to treat
the solution of LB1 and LB2 numerically. We have developed a script
which implements these considerations in Python~\cite{python} and
which is publicly available~\cite{lbweightscode}. It has been written
in such a way that it runs both under Python 2.7 as well as 3.5. The
present paper may therefore also be viewed as the documentation of the
software. The non--trivial aspects of linear algebra are taken care of
by utilizing well--established routines from the
NumPy~\cite{numpy,numpy_linear_algebra} package. As far as we
understand, and which seems to be consensus in the
community~\cite{philippi_continuous_2006}, there is no known method to
find a suitable (smallest) set of lattice velocities with simple
\emph{a priori} criteria; rather one has to choose a set (essentially
by trial and error) and then check if this allows for a solution of
LB2. This is precisely what the script does: It asks the user for
defining a set of shells, and then uses that set for analysis. We
mainly focus on the case where LB2 has \emph{one and only one}
solution (``minimal'' models). This is in spirit quite similar to the
work of Philippi et al.~\cite{philippi_continuous_2006}, and also of
Shan~\cite{shan_pap04_2010, shan_pap05_2016}, however with
significantly reduced mathematical complexity. Those cases where the
problem has no solution whatsoever are obviously discarded. There are
also cases where there are infinitely many solutions. These cases are
not analyzed in a comprehensive fashion, but only by reduction to a
special case, where $\cs$ is given. From there, a unique set of
weights is determined by solving a linear programming problem which
aims at the minimization of some particular weight, or even several of
them. The script is able to treat arbitrary spacial dimensions, and an
arbitrary maximum tensor rank.

At this point, we would like to emphasize that of course a large
fraction of what has been presented so far, and will be presented in
the following sections, is not new. The central importance of
Eq.~\ref{eq:MBC} has been appreciated by numerous authors, and a
significant fraction of them refers to it not in terms of MBCs but
rather in terms of Gaussian integration --- while the mathematical
problem as such is of course identical, regardless of
nomenclature. Secondly, the underlying linear structure of the
problem, and last not least its relation to linear programming, is
also well--known, and has, most notably, been exploited previously in
the work by X. Shan~\cite{shan_pap04_2010, shan_pap05_2016}. As far as
we are aware, Ref.~\onlinecite{shan_pap05_2016} is so far the most
extensive study on the problem, with models that are isotropic up to
tensor rank eight. What is new about our work is (i) the
implementation in terms of publicly available software, (ii) a novel
approach to re--cast the tensor equations in terms of scalar equations
by contraction with random tensors (see Sec.~\ref{sec:algorithm}), and
(iii) the systematic application of numerical linear algebra, without
any complicated group theory. Beyond a perfect reproduction of the
results of Ref.~\onlinecite{shan_pap05_2016}, see
Appendix~\ref{sec:ComparisonShan}, we are also able to find models
with a yet higher degree of isotropy up to tensor rank ten.

The remainder of this paper is organized as follows: The following
section (Sec.~\ref{sec:algorithm}) is devoted to a detailed derivation
and description of the algorithm that has been
implemented. Section~\ref{sec:examples} then demonstrates, via a few
examples, what kind of results can be obtained with the software very
easily. After that, Sec.~\ref{sec:summary} provides a brief summary.

Appendix~\ref{sec:equilpopul} briefly discusses how the obtained
models can be used to construct the equilibrium populations for
nonvanishing flow velocities, using either the Hermite--polynomial
expansion or the entropic approach, which are demonstrated to be
asymptotically equivalent in the limit of full isotropy. This part
does not present new results but is rather intended as background
information to complete the picture; experienced readers can probably
skip that part. Appendix~\ref{sec:ComparisonShan} provides details on
how we used our software to check the results of
Ref.~\onlinecite{shan_pap05_2016}, and
Appendix~\ref{sec:AccuracyTestMode} some numerical details about the
``test'' mode of our script, where the set of weights is not
calculated but rather checked whether it indeed satisfies the MBCs.

\section{Derivation of the algorithm}
\label{sec:algorithm}

\subsection{Linear algebra}
\label{sec:LinearAlgebra}

Let us consider the central relation
\be
  \label{eq:MBCAgain}
  \sum_i w_i c_{i \alpha} c_{i \beta} \ldots c_{i \gamma} =
  \int d^d \vec v f(\vec v) v_\alpha v_\beta \ldots
  v_\gamma .
\ee
This is a tensor identity for tensors of rank $m$, where $m$ is the
number of $\vec c_i$ factors on the left hand side (lhs), or the
number of $\vec v$ factors on the rhs. For odd $m$, the relation is
trivially satisfied for symmetry reasons. We wish to satisfy the
relation for all $m$ with $m \le M$, where $M$ is a user--supplied
even number. The rank $m = 0$ is just the normalization condition for
the weights. The weight $w_0$, corresponding to the velocity $\vec c_i
= 0$, occurs only in that condition but not the other equations. It is
therefore sufficient to first solve the problem for the weights with
nonzero $\vec c_i$, restricting attention to even $m \ge 2$, and then
adjust $w_0$ at the end in order to satisfy normalization.

If we denote the number of shells (excluding the zero velocity shell)
with $\Ns$, enumerate these shells with an index $s = 1, \ldots, \Ns$,
and take into account that the weights are identical within a shell,
the MBCs can be written as
\be
  \label{eq:MBCMoreDetail}
  \sum_{s = 1}^{\Ns} w_s \sum_{i \in s} 
  c_{i \alpha} c_{i \beta} \ldots c_{i \gamma} =
  \int d^d \vec v f(\vec v) v_\alpha v_\beta \ldots
  v_\gamma ,
\ee
to be satisfied for tensor ranks $m = 2, 4, \ldots, M$.

We note that on both sides the tensors are obviously fully symmetric
under arbitrary exchange of indexes. This property alone reduces the
complexity (or dimensionality) of the problem enormously. However, a
further reduction occurs because of geometric symmetry. The rhs is
clearly invariant under reflection, and any rotation in continuous
space, while the lhs is invariant under the cubic group. For the time
being, we view $\cs^2$ as a fixed (``user--supplied'') number, and
therefore we may consider the integrals on the rhs as evaluated, such
that the rhs is simply a known numerical tensor.

We now consider a tensor as a vector in tensor product space. From
symmetry (see also Ref.~\cite{chen_discrete_2008}), we know that both
sides can be expanded in terms of elementary tensors as follows:
\begin{itemize}
\item $m = 2$:
  \bea
  \text{rhs} & = & \ldots \delta_{\alpha \beta} \\
  \text{lhs} & = & \ldots \delta_{\alpha \beta} ;
  \eea
\item $m = 4$:
  \bea
  \text{rhs} & = & \ldots \left( \delta_{\alpha \beta} \delta_{\gamma \delta}
    + \delta_{\alpha \gamma} \delta_{\beta \delta}
    + \delta_{\alpha \delta} \delta_{\beta \gamma} \right) \nonumber \\
    & \equiv & \ldots \left( \delta_{\alpha \beta} \delta_{\gamma \delta}
    + \text{perm.} \right) 
    \\
  \text{lhs} & = & \ldots \left( \delta_{\alpha \beta} \delta_{\gamma \delta}
    + \text{perm.} \right) + \ldots \delta_{\alpha \beta \gamma \delta} ;
  \eea
\item $m = 6$:
  \bea
  \text{rhs} & = & \ldots \left( \delta_{\alpha \beta} \delta_{\gamma \delta} \delta_{\sigma \tau}
    + \text{perm.} \right) \\
  \text{lhs} & = & \ldots \left( \delta_{\alpha \beta} \delta_{\gamma \delta} \delta_{\sigma \tau}
    + \text{perm.} \right) \nonumber \\
  & + & \ldots \left( \delta_{\alpha \beta \gamma \delta} \delta_{\sigma \tau}
      + \text{perm.} \right) \nonumber \\
  & + & \ldots \delta_{\alpha \beta \gamma \delta \sigma \tau}
  \eea
\end{itemize}  
and so on. Here the $\delta$ tensors are generalized Kronecker
symbols, which are one if all indexes are the same and zero
otherwise. The symbol ``perm.'' indicates a suitable set of index
permutations such that the expression under consideration is properly
symmetrized (like explicitly indicated for $m = 4$). The prefactors
``$\ldots$'' are the coefficients which may in principle be calculated
by evaluating Gaussian integrals for the rhs, or lattice sums for the
lhs. We may then consider the tensors $\delta_{\alpha\beta}$,
$\delta_{\alpha \beta} \delta_{\gamma \delta} + \text{perm.}$,
$\delta_{\alpha \beta \gamma \delta}$, etc. as \emph{basis vectors} in
tensor space and the coefficients ``$\ldots$'' as \emph{vector
  components}. From this, we see that the rhs is always an
element of a one--dimensional space, while the dimensionality
of the space corresponding to the lhs depends on the tensor rank
$m$: For $m = 2$, we get a one--dimensional space, for $m = 4$
a two--dimensional space, for $m = 6$ a three--dimensional space,
and so on. 

To discuss the ``and so on'' in more detail, let us first introduce a
short--hand notation and simply write $(2)$ for a second--rank
Kronecker tensor, $(2,2)$ for the symmetrized product of two
second--rank Kronecker tensors, $(4)$ for a fourth--rank Kronecker
tensor, etc.. We may then say that the space for $m = 2$ has the basis
$(2)$, while $m = 4$ has the basis $(4), (2,2)$, and $m = 6$ has the
basis $(6), (4,2), (2,2,2)$. For $m = 8$ we then get
$(8), (6,2), (4,4), (4,2,2), (2,2,2,2)$ or a five--dimensional space.
This process continues: For each higher $m$, we get a new tensor
$(m)$, plus all possible products of the lower--order tensors. In
general, we thus get a tensor space dimension $D_T(m)$ for $m$th rank
tensors, and this may be calculated easily in Python by explicitly
constructing the patterns $(m), (m, m - 2), \ldots$ from the
lower--order patterns in a recursive fashion. As far as we understand,
there is no closed formula for $D_T(m)$; in number theory, $D_T(m)$ is
known as the ``partition function'' (or ``number of partitions'') of
$m/2$ (see e.~g. Ref.~\cite{wikipedia_partition_2018}). For given $M$,
the script therefore calculates (and stores) the dimensions $D_T(m)$
for all $m = 2, 4, \ldots, M$, as well as the dimension of the total
space (comprising tensors of all the ranks under consideration), which
is
\be
\label{eq:defineR}
R = \sum_{m = 2, 4, \ldots}^M D_T(m) .
\ee
It is also clear that, for each $m$, not only the lhs but also the rhs
of Eq.~\ref{eq:MBCMoreDetail} must be an element of the
$D_T(m)$--dimensional symmetry--restricted subspace, since the cubic
group is a subgroup of the full rotation--and--reflection group of
continuous space.

The problem, however, is that this consideration yields only the
\emph{maximum} dimension of the subspace of all the tensors whose form
is that of the lhs. The number $D_T(m)$ is just a consequence of
symmetry, while the actual dimension is a result of the supplied
velocity set: The true subspace is the span of the elementary tensors
$\sum_{i \in s} c_{i \alpha} c_{i \beta} \ldots c_{i \gamma}$, and
this may, for a poorly chosen (or simply too small) set, be
\emph{smaller} than the space of tensors that are symmetric with
respect to the cubic group, and to the permutation group of the
indexes. In that situation it may actually occur that the rhs is not
an element of that smaller space, or, in other words, that there is no
set of weights that solves Eq.~\ref{eq:MBCMoreDetail}. Conversely, it
may also turn out that the velocity set is chosen rather large, such
that the equations have infinitely many solutions. An important aspect
of the software is therefore that it has to be able to reliably detect
such cases.

In the present paper, we propose to start from Eq.~\ref{eq:MBCMoreDetail}
and to contract it with an elementary tensor of rank $m$
\be
  \label{eq:ElementaryTensor}
  n_\alpha n_\beta \ldots n_\gamma ,
\ee
where $\vec n$ is some unit vector
($\left\vert \vec n \right\vert = 1$), chosen with \emph{random}
orientation, uniformly distributed on the $d$--dimensional sphere. In
this way, we project the tensor equation onto a scalar equation.  In
this context, it should be recalled that contraction over all indexes
of two tensors of the same rank naturally defines a scalar product in
tensor product space, which then immediately allows one to construct
the geometric concept of an orthogonal projection. It should also be
noted that the elementary tensors are invariant under index
permutation but \emph{not} under any geometric symmetry
transformation.

We do this contraction not only for one unit vector but for $D_T(m)$
unit vectors for the tensor equation of rank $m$, and do this for all
ranks $m = 2, 4, \ldots, M$. We thus obtain $R$ scalar equations, and
for each of these equations we generate a new unit vector $\vec n_r$,
$r = 1, \ldots, R$. Let us denote the rank corresponding to the $r$th
equation with $m_r$.

On the rhs we then obtain~\cite{gaussian_2017}
\begin{eqnarray}
  \nonumber
  && \int d^d \vec v \, f(\vec v)
  \left(\vec v \cdot \vec n_r \right)^{m_r} \\
  \nonumber
  & = & (2 \pi \cs^2)^{-1/2} \int_{- \infty}^{+\infty} 
  dv_x \, v_x^{m_r} \exp\left( - \frac{v_x^2}{2 \cs^2} \right) \\
  & = & (m_r - 1)!! \, \cs^{m_r} ,
\end{eqnarray}
where $(m - 1)!! = (m - 1) (m - 3) (m - 5) \ldots 3 \cdot 1$.

We therefore define
\be 
  A_{rs} = \frac{1}{(m_r - 1)!!} \sum_{i \in s} 
  \left( \vec c_i \cdot \vec n_r \right)^{m_r} , 
\ee 
which can be straightforwardly calculated as soon as the velocity
shells are specified and the random vectors are generated. Then the
resulting set can be written as
\be 
  \label{eq:FundamentalSystem}
  \sum_{s = 1}^{\Ns} A_{rs} \, w_s = \cs^{m_r} ,
\ee
which is obviously a set of linear equations to determine the weights
$w_s$. In matrix form this is written as
\be
  \label{eq:MatrixEquation}
  A \vec w = \vec b ,
\ee
where $A$ is the $R \times \Ns$ matrix formed by the elements
$A_{rs}$, $\vec w$ the vector of weights, and $\vec b$ the rhs vector
according to Eq.~\ref{eq:FundamentalSystem}. Our strategy is thus to
construct this set of equations and to solve it numerically.

Let us now discuss why we believe that this procedure is correct and
useful. Within a given tensor rank $m$, we have $D_T(m)$ elementary
tensors $n_\alpha n_\beta \ldots n_\gamma$. It is then highly probable
that these tensors are all linearly independent. Actually, in our
opinion this is much more probable than linear independence of a set
of elementary tensors chosen by a guessing and erring human. More
importantly, though, it is highly likely that the \emph{projections}
of the elementary tensors onto the $D_T(m)$--dimensional subspace of
invariant tensors are still linearly independent. If that is the case,
then the contractions, i.~e. the scalar products of the elementary
tensors with the lhs tensor, provide enough information in order to
characterize the latter \emph{uniquely}. In other words: Our
thus--generated $R$ scalar equations are \emph{equivalent} to the
original set of tensor equations.

The easiest case occurs obviously when $A$ is quadratic ($\Ns = R$)
and non--degenerate, because then Eq.~\ref{eq:MatrixEquation} can be
solved by simple inversion. Therefore the script first calculates $R$
and then suggests to pick precisely $R$ shells --- but the user has
the freedom to follow that suggestion or not; i.~e. both $\Ns > R$ as
well as $\Ns < R$ are permitted. Typically, one expects infinitely
many solutions for $\Ns > R$ and no solution whatsoever for $\Ns < R$;
however, due to degeneracies this does not always have to be the
case. Similarly, picking $\Ns = R$ does not guarantee at all that $A$
is non--degenerate. A significant part of the software therefore aims
at treating these less straightforward cases.

At this point, it is useful to consider $\cs^2$ no longer as a fixed
number but rather as a parameter that can be varied. Since the rhs
$\vec b$ consists of $\cs^2$, $\cs^4$, $\ldots$, $\cs^M$, it is clear
that the weights must be polynomials in $\cs^2$. Therefore we write
\be
  \label{eq:PolynomialForWs}
  w_s = \sum_{\mu = 2, 4, \ldots}^M q_{s \mu} \, \cs^\mu ,
\ee
resulting in
\be
  \sum_\mu \cs^\mu \sum_{s} A_{rs} \, q_{s \mu} = \cs^{m_r}
  = \sum_\mu \cs^\mu \delta_{m_r \mu} .
\ee
Comparing coefficients, we find
\be
  \label{eq:QmatrixEquationElements}
  \sum_{s} A_{rs} \, q_{s \mu} = \delta_{m_r \mu} =: D_{r \mu}
\ee
or, in matrix form
\be
  \label{eq:QmatrixEquation}
  A Q = D .
\ee
Our aim is therefore to solve that system to calculate the matrix $Q$,
such that we find a solution that is valid for \emph{any} possible
value of $\cs$ --- note that Eq.~\ref{eq:QmatrixEquation} no longer
contains $\cs^2$.

The first step of the analysis is a standard singular--value
decomposition~\cite{press_numerical_2007, singular_2017}, which is
possible for any matrix $A$ independent of its shape or rank. The
NumPy package provides a routine to do this~\cite{numpy_svd}. The
decomposition reads
\be
  \label{eq:oursvd}
  A = U S V^T ,
\ee
where $S$ is a rectangular matrix of the same shape ($R \times \Ns$)
as $A$, and $U$ and $V$ are quadratic orthogonal matrices of suitable
size ($R \times R$ and $\Ns \times \Ns$), with $U^T U = 1$,
$V^T V = 1$ (unit matrices). Here the superscript $T$ denotes
transposition. $S$ is a matrix consisting of all zeros, except the
entries $S_{11} = \sigma_1 > 0$, $S_{22} = \sigma_2 > 0$, $\ldots$,
$S_{ZZ} = \sigma_{Z} > 0$ (the singular values). Here of course it has
to be checked if some ``nonzero'' singular values have only been
produced as a result of numerical roundoff errors. Obviously, $Z \le
\min(\Ns,R)$. $Z$ is the rank of $S$ (or of
$A$), and maximum rank occurs for $Z = \min(\Ns,R)$, while for
$Z < \min(\Ns,R)$ the problem is rank--deficient.

Inserting Eq.~\ref{eq:oursvd} into Eq.~\ref{eq:QmatrixEquation}, one
sees that the problem is equivalent to
\be
  \label{eq:ReducedEquation}
  S Q' = D'
\ee
with the abbreviations
\bea
  Q' & = & V^T Q , \\
  D' & = & U^T D .
\eea
As $Z \le R$, we can only have the cases $Z = R$ or $Z < R$.
Let us first treat the latter case, for which there are
$R - Z$ equations of the form
\bea
&&
\begin{pmatrix}
  0      & \dots & 0      \\
  \vdots &       & \vdots \\
  0      & \dots & 0
\end{pmatrix}
\\
\nonumber
& = &
\begin{pmatrix}
  D'_{r=Z+1,\mu=2} & D'_{r=Z+1,\mu=4} & \dots & D'_{r=Z+1,\mu=M} \\
  \vdots           & \dots             & \dots & \vdots          \\
  D'_{r=R,\mu=2}   & D'_{r=R,\mu=4}   & \dots & D'_{r=R,\mu=M}
\end{pmatrix} .
\eea
This can obviously only hold if the rhs vanishes, and this can be
easily checked by calculating the Frobenius norm of the latter, using
the standard NumPy routine ``norm''~\cite{numpy_norm}. This is nothing
but the criterion for the existence of a solution, and if it fails,
the script aborts, and informs the user. This situation means that the
set of shells is either too small or chosen inappropriately, such that
degeneracies occur. The user is then encouraged to try again with a
different set of shells.

Conversely, if the check succeeds, then the equations number $Z + 1$,
$Z+ 2$, \ldots, $R$ may simply be discarded. Doing this, we arrive at
a simplified matrix $\tilde S$ of size $Z \times \Ns$, as well as a
simplified rhs $\tilde D'$.

If $Z = R$, no such ``pruning'' needs to be done, and we simply have
$\tilde S = S$, $\tilde D' = D'$. We thus arrive at a simplified set
\be
  \label{eq:SimplifiedEquation}
  \tilde S Q' = \tilde D' .
\ee
As a next step, we scale the equations by $1/\sigma_1$, $1/\sigma_2$,
\ldots, $1/\sigma_Z$, resulting in
\be
  \label{eq:EvenSimplerEquation}
  \tilde S' Q' = \tilde D'' .
\ee
$\tilde S'$ is a trivial matrix whose nonzero entries are all one.
Now, since $\Ns \ge Z$, the matrix $\tilde S'$ can either be quadratic
($\Ns = Z$) or rectangular, with more columns than rows ($\Ns > Z$).
In the former case, $\tilde S'$ is simply the unit matrix, such that
the solution is unique and directly found via $Q' = \tilde D''$ or
$Q = V \tilde D''$, from which the weights are found as polynomials
in $\cs^2$, returned, and further processed according to subsections
\ref{sec:RangeOfValidity} and \ref{sec:RationalNumbers}.

For $\Ns > Z$ we have infinitely many solutions. To treat this latter
case, we also provide some numerical procedures, however in a less
comprehensive and ambitious fashion. The matrices $V$ and $\tilde D''$
(from which $\tilde S'$ can be easily re--constructed), together with
necessary information about the shells, are stored in a file, which is
then processed further in a separate script ``Continue.py''. This will
be the topic of subsection \ref{sec:RankDeficiency}.

\subsection{Range of validity}
\label{sec:RangeOfValidity}

Assuming that the script has found a unique solution by making use of
linear algebra, we still have not yet satisfied one important
condition: For physical reasons, the populations $n_i$ should be
positive, which in turn means that the weights $w_s$ must be positive
as well. This is however typically only true within one (or more)
narrow interval(s) of $\cs^2$ values. It may even turn out that there
is no $\cs$ value whatsoever that satisfies the condition. It is
therefore desirable that the script automatically finds this range of
validity.  This is facilitated by the NumPy routine
``roots''~\cite{numpy_roots}, which returns all complex roots of a
polynomial given in terms of its coefficients. This procedure is
applied to all the functions $w_s(\cs^2)$ that the linear algebra
routines have found. Technically, ``roots'' is a linear algebra
routine as well, since it is based upon mapping the root--finding
problem onto an eigenvalue problem.

The script then eliminates all roots with non--vanishing imaginary
part, as well as all roots with real part $\le 0$. The remaining $K$
roots $z_1, z_2, \ldots, z_K$ are arranged in a sorted array, making
use of the NumPy routine ``sort''~\cite{numpy_sort}. This defines a
sequence of intervals $(0, z_1)$, $(z_1, z_2)$, \ldots, $(z_{K - 1},
z_K)$, $(z_K, \infty)$, in which no change of sign can occur. By
evaluating all functions $w_s(\cs^2)$ in the centers of these
intervals (i.~e. at the points $(z_{n + 1} + z_n) / 2$), we can
eliminate all the intervals that violate the condition of positivity
of weights. For the last interval, the functions are evaluated at
$(3/2) z_K$. Typically --- but not always --- this procedure finds one
single interval of validity.

The special $\cs^2$ values that form the limits of validity (``magic''
$\cs^2$ values) are characterized by the vanishing of at least one
weight $w_s$. In this case, the corresponding shell(s) can be
discarded completely, which gives rise to a ``reduced'' model, which
is often useful in practice. For this reason, the script evaluates all
weights at the ``magic'' $\cs^2$ values, such that the user receives
quick information about the properties of the resulting reduced
models.

\subsection{Velocity shells}
\label{sec:VelocityShells}

We recall that a shell is defined as an equivalence class of lattice
speeds that can be mapped onto each other by an element of the cubic
point group of the lattice. A less strict concept is that of a
``modulus shell'' that comprises all lattice speeds whose modulus is
the same. In general, a modulus shell can be decomposed into several
subshells, each of which is an equivalence class of its own. Therefore
the script proceeds in several steps in order to define the shells:
(i) Construction of the cubic group in $d$ dimensions, (ii) supply of
modulus values by the user, (iii) finding the corresponding modulus
shells, (iv) decomposing the modulus shells into subshells, and (v)
possible \emph{a posteriori} elimination of some of the thus--found
shells by the user.

Let us first discuss the construction of the cubic group in $d$
dimensions. Denoting the Cartesian unit (column) vectors with $\vec
e_1$, $\vec e_2$, \ldots, $\vec e_d$, we see that the $d \times d$
unit matrix is written as $(\vec e_1, \vec e_2, \ldots, \vec e_d)$. A
transformation that is just tantamount to a permutation $\pi$ of the
Cartesian axes therefore corresponds to the matrix $(\vec e_{\pi(1)},
\vec e_{\pi(2)}, \ldots, \vec e_{\pi(d)})$.  Combining this with the
possibility to flip the orientation of an axis, the most general
transformation matrix of the cubic group has the form $(\pm \vec
e_{\pi(1)}, \pm \vec e_{\pi(2)}, \ldots, \pm \vec e_{\pi(d)})$, where
each combination of signs is possible. Based upon these observations,
it is very easy to construct the set of all transformation matrices,
whose number therefore turns out to be $d! 2^d$ (i.~e. eight in two
dimensions, $48$ in three dimensions). We here make use of the
``permutations'' routine of the ``itertools'' section of the standard
Python library, plus the observation that any sign combination can be
written as a string of pluses and minuses. Such a string is
straightforwardly mapped onto a corresponding string of zeros and
ones. Such a string, in turn, is identified with the binary
representation of an integer in the range $0, 1, \ldots, 2^d - 1$,
which therefore just needs to be scanned in order to find all sign
combinations.

In the next step, the user specifies the squared moduli of the desired
velocities. For one modulus shell, we thus have an integer number
$L = \vec c_i^2$. The corresponding vectors are then being searched
for by the script. Obviously, it is sufficient to search a
$d$--dimensional cubic grid, where each coordinate varies from $-L$ to
$+L$. The total number of points to be scanned is thus $(2 L +
1)^d$. Introducing a trivial coordinate shift, one may as well search
a grid whose coordinates vary from $0$ to $2 L$. A corresponding
one--dimensional index $k$ that scans all grid points then varies from
$0$ to $(2 L + 1)^d - 1$. This index is related to the shifted
coordinates $x_1, x_2, \ldots, x_d$ via
\be
  k = \sum_{l = 1}^d (2 L + 1)^{l - 1} \, x_l .
\ee
Therefore these coordinates can be retrieved from $k$ recursively by
successive modulo operations. After having collected and shifted the
thus--found coordinates, the program calculates the squared modulus
and checks if that value coincides with $L$. If yes, the vector is
added to a list.

The two final steps (iv) and (v) are then straightforward and
need not be explained in further detail.

\subsection{Random vectors}

Using a uniform random generator (the script uses the built-in
generator that is provided by Python via the ``random'' package), it
is very easy to generate $d$ coordinates $x_i$ distributed uniformly
in the interval $(-1, 1)$. We then calculate $\sum_i x_i^2$ and check
if this is smaller than one. If not, the procedure is repeated until
the criterion is satisfied. The thus--found vector $\vec x$ is then
normalized to unity, yielding $\vec n = \vec x / \left\vert \vec x
\right\vert$. It is clear that the thus--generated vector $\vec n$ is
a unit vector that is uniformly distributed on the unit sphere.

\subsection{Rational numbers}
\label{sec:RationalNumbers}

Considering the expansion of $w_s$ in powers of $\cs^2$
(Eq.~\ref{eq:PolynomialForWs}), and the original equations in the form
of Eq.~\ref{eq:tensor0} to \ref{eq:tensor4}, it is quite clear that
the coefficients $q_{s \mu}$ can be viewed as the solution to a system
of linear equations whose coefficients are all integer. For this
reason, they must be simple rational numbers. Since, e.~g., a fraction
like $1 / 24$ is more intuitive and aesthetically more appealing than
the corresponding floating--point number $0.04166666$, the script
makes use of a routine that converts the latter into the former. In
principle, this is done via a standard continued--fraction
expansion~\cite{continued_2017}, which is however somewhat tricky to
implement due to its high sensitivity to roundoff errors. Fortunately,
Python provides the ready--made routine
``Fraction''~\cite{python_fractions} which yields quite satisfactory
results if the size of the denominator is suitably limited, and
the model is of sufficiently low order, such that the denominators
are not too large.

This conversion is also applied to the ``magic'' $\cs^2$ values and to
the coefficients of the resulting reduced models. However, these might
be irrational, in which case the procedure provides fractions with
large numerators and denominators. If the user is interested in exact
algebraic numbers, we recommend to identify the algebraic equation
whose solution provides the magic $\cs^2$ value, and to attempt its
exact solution with the help of a computer algebra system such as
Wolfram Alpha~\cite{wolframalpha}. It should be stressed, though, that
for practical purposes a floating--point representation is absolutely
sufficient.

\subsection{The case of infinitely many solutions}
\label{sec:RankDeficiency}

For rank--deficient problems that have infinitely many solutions, we
do not attempt to find the weights as a function of $\cs^2$, but
rather only for one particular $\cs^2$ value, for which the user is
explicitly asked. We do this in a separate script ``Continue.py'',
which obtains its further input from a file written by the main
script.

Starting from Eq.~\ref{eq:EvenSimplerEquation}, which we write in the
form
\be
  \label{eq:EvenSimplerEquationNewVersion}
  \tilde S' V^T Q = \tilde D'' ,
\ee
and recalling that the sought--for matrix $Q$ contains the
coefficients of the polynomial expansions of the weights $w_s$, we see
that we can, for a given (user--supplied) $\cs^2$ value, immediately
construct a set of linear equations that the weights have to satisfy.
We know that this set has infinitely many solutions. Furthermore, we
know that all weights have to satisfy the conditions $w_s \ge 0$. If
we then combine this with some linear optimization problem, we see
that this is identical to a problem of standard linear
programming~\cite{schrijver_theory_1998}. The most useful optimization
problem that we can imagine in this context is to minimize one of the
weights, or perhaps even several of them. The user is therefore asked
which of the weights is to be minimized; in case that several weights
are supplied, the script simply attempts to minimize the sum of these
weights.

For our purposes, we found the package ``cvxpy''~\cite{cvxpy_ref}
particularly useful in terms of (i) Python integration, (ii)
correctness of results, and (iii) numerical stability. The script
checks if the problem has a solution, and if yes, it returns it,
together with the $\cs^2$ value. Quite often, the minimized weight
turns out to be zero.  To enhance the ease of use, the user may supply
a whole interval of $\cs^2$ values plus a step size, such that the
whole interval is being scanned.

In practical applications, it often turns out that it is useful to
first supply a fairly large set of shells, which then results in a
rank--deficient problem, and to then use ``Continue.py'' to remove
more and more shells until finally a minimal model is found.

\subsection{The ``test'' mode}
\label{sec:test_mode}

Except for solving the problem of finding weights from scratch, quite
frequently the situation arises where one is confronted with a given
(or claimed) solution (e.~g. from the literature), and one would like
to quickly check its correctness. The script therefore provides a
``test'' mode, where the formalism developed above is used for that
purpose. Input data are therefore not only spacial dimension, maximum
tensor rank, and the set of shells (as always), but additionally the
value of $\cs$, plus the set of weights that should be tested. Note
that in ``test'' mode the script assumes that a given solution has
been given for one special well--defined $\cs$ value, and also
disregards the problem of positivity of weights. Therefore, the task
is to simply check if the given vector of weights $\vec w$ satisfies
Eq.~\ref{eq:MatrixEquation}, which is easy, because the provided
information allows to calculate both the matrix $A$ and the
inhomogeneity $\vec b$. In case that the given solution is provided
simply as a set of numbers (a vector $\vec w_0$), we therefore
calculate the residual
\be
\vec \Delta_0 = A \vec w_0 - \vec b ,
\ee
and analyze whether it is zero within numerical accuracy. In the
more general case of a degenerate solution, we assume that it is
given in the form
\be
\label{eq:ParametricSolution}
\vec w = \vec w_0 + \sum_{i \ge 1} \lambda_i \vec w_i ,
\ee
where the $\lambda_i$ form a set of parameters which may be varied
independently. Obviously, we again have to evaluate $\vec \Delta_0$ as
before, and check for its vanishing, but additionally we also need to
evaluate the additional residuals $\vec \Delta_i = A \vec w_i$,
$i \ge 1$, and check for their vanishing as well.

Given the fact that literature values for weights are typically given
with not more than six--digit accuracy, we need to take care that the
check for vanishing residuals is not too stringent. How this is done
in detail is explained in Appendix~\ref{sec:AccuracyTestMode}.

\subsection{The algorithm as a whole}

\begin{figure}
  \includegraphics[width=0.7\textwidth]{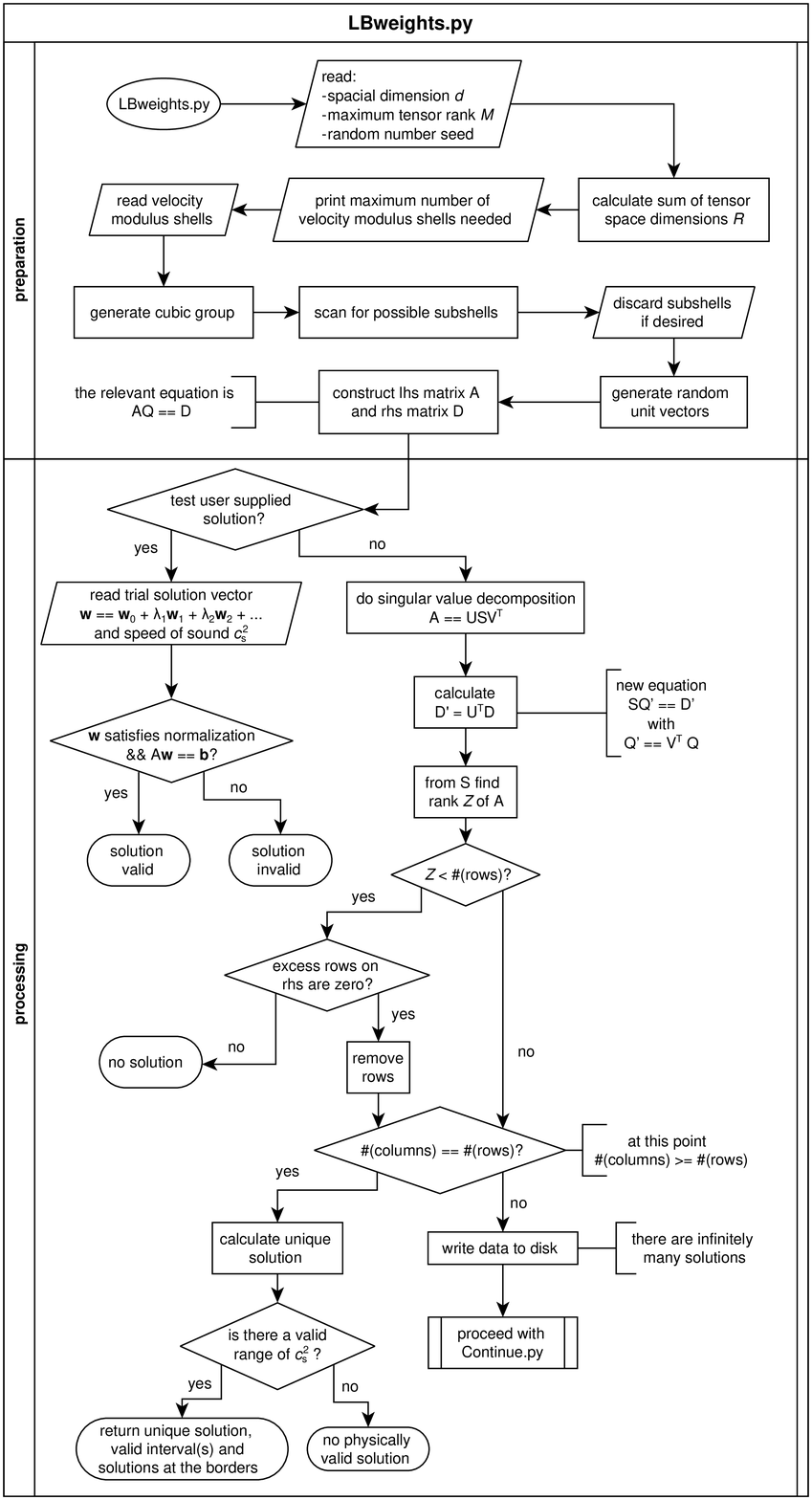}
    \caption{Flow diagram of the algorithm.}
  \label{fig:FlowDiagram}
\end{figure}

The considerations given above give rise to a procedure which is
summarized in the flow diagram Fig.~\ref{fig:FlowDiagram}.
In general, input data may be provided either by an interactive
dialogue or via command--line arguments.

\onecolumngrid

\twocolumngrid

\clearpage

\section{Examples}
\label{sec:examples}

\subsection{Two--dimensional models}

We start with maximum tensor rank $M = 4$, i.~e. $R = 3$, such that
one expects that three non--trivial shells are necessary.  Indeed
attempts to solve the problem with one or two such shells turned out
to be unsuccessful. Trying the three shells $c_i^2 = 1, 2, 4$ (with
typical vectors $(1,0)$, $(1,1)$ and $(2,0)$, respectively, such that
in total one has $13$ vectors), yields the solution
\begin{eqnarray}
w_{(00)} & = & 1 - \frac{5}{2} \cs^2 + \frac{5}{2} \cs^4 \\
w_{(10)} & = & \frac{2}{3} \cs^2 - \cs^4 \\
w_{(11)} & = & \frac{1}{4} \cs^4 \\
w_{(20)} & = & - \frac{1}{24} \cs^2 + \frac{1}{8} \cs^4 ,
\end{eqnarray}
which is valid in the interval $1/3 \le \cs^2 \le 2/3$. For $\cs^2 =
1/3$ one obtains a reduced model with nine velocities and weights
$w_{(00)} = 4/9$, $w_{(10)} = 1/9$, $w_{(11)} = 1/36$; this is nothing
but the well--known D2Q9 model~\cite{qian_lattice_1992}. Another
nine--velocity model is obtained for $\cs^2 = 2/3$ with $w_{(00)} =
4/9$, $w_{(11)} = 1/9$, $w_{(20)} = 1/36$.

We continue with $M = 6$, i.~e. $R = 6$. Attempting a six--shell
model with $c_i^2 = 1, 2, 4, 5, 8, 9$ gives rise to a rank--deficient
problem with infinitely many solutions. Removing the shell $c_i^2 = 5$
(it is the most natural candidate since it contains most velocities)
yields indeed a unique solution given by
\begin{eqnarray}
  w_{(00)} & = & 1 - \frac{49}{18} \cs^2
  + \frac{175}{48} \cs^4 - \frac{85}{48} \cs^6 \\
  w_{(10)} & = & \frac{3}{4} \cs^2 - \frac{71}{48} \cs^4
  + \frac{13}{16} \cs^6 \\
  w_{(11)} & = & \frac{1}{3} \cs^4 - \frac{1}{4} \cs^6 \\
  w_{(20)} & = & - \frac{3}{40} \cs^2 + \frac{25}{96} \cs^4
  - \frac{5}{32} \cs^6 \\
  w_{(22)} & = & - \frac{1}{192} \cs^4 + \frac{1}{64} \cs^6 \\
  w_{(30)} & = & \frac{1}{180} \cs^2 - \frac{1}{48} \cs^4
  + \frac{1}{48} \cs^6 .
\end{eqnarray}
Here and in what follows the subscripts denote the typical vectors
corresponding to each shell. The model comprises in total 21
velocities (four velocities in each non--trivial shell), and its range
of validity is $0.3702519 \le \cs^2 \le 1.148412$ (irrational
numbers). The lower boundary is the root of $w_{(20)}$ and given by
the exact value $5/6 - \sqrt{193}/30$. The reduced model is thus a
$17$--velocity model with (irrational) weights $w_{(00)} = 0.4020051$,
$w_{(10)} = 0.1161549$, $w_{(11)} = 0.03300635$, $w_{(22)} = 7.907860
\times 10^{-5}$, and $w_{(30)} = 2.584145 \times 10^{-4}$. The upper
boundary, $\cs^2 = 1.148412$,is the root of $w_{(00)}$; this value can
still be given as an exact but unwieldy number. The reduced model in
this case comprises $20$ velocities with weights $w_{(10)} =
0.1411090$, $w_{(11)} = 0.06097080$, $w_{(20)} = 0.02066598$,
$w_{(22)} = 0.01679637$, $w_{(30)} = 0.01045786$.

Another solution is obtained if the last shell ($c_i^2 = 9$) is
replaced by $c_i^2 = 16$ (also four vectors):
\begin{eqnarray}
  w_{(00)} & = & 1 - \frac{21}{8} \cs^2
  + \frac{105}{32} \cs^4 - \frac{45}{32} \cs^6 \\
  w_{(10)} & = & \frac{32}{45} \cs^2 - \frac{4}{3} \cs^4
  + \frac{2}{3} \cs^6 \\
  w_{(11)} & = & \frac{1}{3} \cs^4 - \frac{1}{4} \cs^6 \\
  w_{(20)} & = & - \frac{1}{18} \cs^2 + \frac{3}{16} \cs^4
  - \frac{1}{12} \cs^6 \\
  w_{(22)} & = & - \frac{1}{192} \cs^4 + \frac{1}{64} \cs^6 \\
  w_{(40)} & = & \frac{1}{1440} \cs^2 - \frac{1}{384} \cs^4
  + \frac{1}{384} \cs^6 ;
\end{eqnarray}
this model is valid for $0.3510760 \le \cs^2 \le 4/3$. The former
value is again an irrational number given by the root of $w_{(20)}$;
its exact value is $9/8 - \sqrt{115/192}$. The resulting reduced
17--speed model at $\cs^2 = 0.3510760$ is given by the weights
$w_{(00)} = 0.4220031$, $w_{(10)} = 0.1141627$, $w_{(11)} =
0.03026688$, $w_{(22)} = 3.416974 \times 10^{-5}$, $w_{(40)} =
3.551447 \times 10^{-5}$. At the upper limit $\cs^2 = 4/3$ the reduced
model comprises only 16 speeds, since at $\cs^2 = 4/3$ both $w_{(00)}$
and $w_{(11)}$ vanish. The remaining weights in this case are
$w_{(10)} = 64/405$, $w_{(20)} = 5/81$, $w_{(22)} = 1/36$, $w_{(40)} =
1/405$.

We now turn to $M = 8$, i.~e. $R = 11$. We thus first attempted the
set $c_i^2 = 1, 2, 4, 5, 8, 9, 10, 13, 16, 18, 25$. The shell $c_i^2 =
25$ comprises two subshells (with vectors of types $(5,0)$ and
$(4,3)$, respectively), such that the set actually gives rise to a
twelve--speed model. Not surprisingly, this results in a
rank--deficient problem with infinitely many solutions. However, the
rank turns out to be merely eight, which indicates that it might be
possible to reduce the model to just eight non--trivial shells. We
therefore tried by removing the outer shells $c_i^2 = 25, 18, 16$;
this however gives rise to a problem with no solution whatsoever.
Excluding $c_i^2 = 13$ instead of $c_i^2 = 16$ gives a set $c_i^2 = 1,
2, 4, 5, 8, 9, 10, 16$, which then indeed provides a unique
solution. Each non--trivial shell comprises four vectors except $c_i^2
= 5, 10$, which contain eight vectors each. All in all, this is
therefore a $41$--speed model. The weights are given by
\begin{eqnarray}
  w_{(00)} & = & 1 - \frac{205}{72} \cs^2 + \frac{1333}{288} \cs^4
  - \frac{205}{48} \cs^6 + \frac{169}{96} \cs^8 \\
  w_{(10)} & = & \frac{4}{5} \cs^2 - \frac{179}{90} \cs^4
  + \frac{9}{4} \cs^6 - \frac{25}{24} \cs^8 \\
  w_{(11)} & = & \frac{19}{36} \cs^4 - \frac{47}{48} \cs^6
  + \frac{9}{16} \cs^8 \\
  w_{(20)} & = & - \frac{1}{10} \cs^2 + \frac{7}{16} \cs^4
  - \frac{7}{12} \cs^6 + \frac{7}{24} \cs^8 \\
  w_{(21)} & = & - \frac{2}{45} \cs^4 + \frac{1}{6} \cs^6
  - \frac{1}{8} \cs^8 \\
  w_{(22)} & = & \frac{1}{576} \cs^4 - \frac{1}{96} \cs^6
  + \frac{1}{64} \cs^8 \\
  w_{(30)} & = & \frac{4}{315} \cs^2 - \frac{1}{18} \cs^4
  + \frac{1}{12} \cs^6 - \frac{1}{24} \cs^8 \\
  w_{(31)} & = & \frac{1}{360} \cs^4 - \frac{1}{96} \cs^6
  + \frac{1}{96} \cs^8 \\
  w_{(40)} & = & - \frac{1}{1120} \cs^2 + \frac{7}{1920} \cs^4
  - \frac{1}{192} \cs^6 + \frac{1}{384} \cs^8 .
\end{eqnarray}
The range of validity is $0.6979533 \le \cs^2 \le 0.8704738$; these
numbers are the irrational roots of $w_{(40)}$ and
$w_{(30)}$. Removing the corresponding shells then gives rise to two
$37$--speed models. At the lower bound we thus obtain the weights
$w_{(00)} = 0.2331507$, $w_{(10)} = 0.1073061$, $w_{(11)} =
0.05766786$, $w_{(20)} = 0.01420822$, $w_{(21)} = 0.005353049$,
$w_{(22)} = 0.001011938$, $w_{(30)} = 2.453010 \times 10^{-4}$,
$w_{(31)} = 2.834143 \times 10^{-4}$. Comparison with
Ref.~\cite{philippi_continuous_2006} shows that this set of velocities
and weights is identical to the model derived by Philippi et al. under
the name ``D2V37'' model.

The two--dimensional models that were investigated in a recent paper
by Shan~\cite{shan_pap05_2016} (going up to tensor order $M = 8$)
could all be verified (except for one minor typo), see
Appendix~\ref{sec:ComparisonShan} .

Furthermore, it is also possible to study the case of tenth--order
isotropy, corresponding to $R = 18$. Starting from the eighteen
velocities
$c_i^2 = 1, 2, 4, 5, 8, 9, 10, 13, 16, 17, 18, 20, 25, 32, 36, 37, 40,
52$, one finds that this yields a rank--deficient problem with
infinitely many solutions, where the rank of the problem is
eleven. Removing outer shells, we can reduce this to the set
$c_i^2 = 1, 2, 4, 5, 8, 9, 10, 13, 16, 25$, which corresponds to
eleven shells (the shell $c_i^2 = 25$ is decomposed into two
subshells, while all others are irreducible). This is a $61$--speed
model with a unique solution and a range of validity of
$0.7592510 \le \cs^2 \le 0.9054850$. We do not give the expansion
of the weights as polynomials in $\cs^2$ here; the expressions
are lengthy and the rational representations of the floating--point
numbers most probably affected by roundoff errors. The other properties
of the model are summarized in Tab.~\ref{tab:d2m10}

\begin{table}
  \begin{tabular}{| c | c | c | c |}
    \hline
    shell & typical & weight at
    & weight at \\
    size  & vector  & $\cs^2 = 7.592510 \times 10^{-1}$
    & $\cs^2 = 9.054850e \times 10^{-1}$ \\
    \hline
    1 & (0, 0) & $2.112895 \times 10^{-1}$ & $1.959760 \times 10^{-1}$ \\
    4 & (0, 1) & $1.069112 \times 10^{-1}$ & $8.636013 \times 10^{-2}$ \\
    4 & (1, 1) & $5.762669 \times 10^{-2}$ & $6.908441 \times 10^{-2}$ \\
    4 & (0, 2) & $1.553262 \times 10^{-2}$ & $2.475221 \times 10^{-2}$ \\
    8 & (1, 2) & $7.296648 \times 10^{-3}$ & $7.207641 \times 10^{-3}$ \\
    4 & (2, 2) & $1.223360 \times 10^{-3}$ & $3.412996 \times 10^{-3}$ \\
    4 & (0, 3) & $5.093571 \times 10^{-4}$ & $4.017308 \times 10^{-4}$ \\
    8 & (1, 3) & $3.635670 \times 10^{-4}$ & $1.260298 \times 10^{-3}$ \\
    8 & (2, 3) & $2.612793 \times 10^{-5}$ & $0                     $ \\
    4 & (0, 4) & $0                     $ & $5.146050 \times 10^{-5}$ \\
    4 & (0, 5) & $8.779627 \times 10^{-7}$ & $6.703596 \times 10^{-7}$ \\
    8 & (3, 4) & $4.044500 \times 10^{-7}$ & $3.253235 \times 10^{-6}$ \\
    \hline
  \end{tabular}
  \caption{Properties of a 61--speed model in two dimensions that
    is isotropic up to tensor rank 10.}
  \label{tab:d2m10}
\end{table}

One thus sees, from the reduced model at the upper limit, that
in two dimensions it is possible to construct a model that is
isotropic up to tenth order with as few as $53$ velocities.

Let us now finally comment on the case of rank--deficient problems
with infinitely many solutions. The main virtue of such models is that
they are able to extend the admissible range of $\cs^2$ values,
however at the expense of more lattice speeds. To illustrate this, let
us again go back to the simple case $M = 4$. As we have seen already,
the set $c_i^2 = 1, 2, 4$ yields a minimal model with range of
validity $1/3 \le \cs^2 \le 2/3$. We now add one further shell
$c_i^2 = 5$ (i.~e. we enhance the model from $13$ speeds to $21$
speeds), which results in a rank--deficient problem, which we analyze
using ``Continue.py'' as described, where we demand that the weight of
the additional shell should be as small as possible. Scanning for
admissible $\cs^2$ values, we find that the lower bound remains
unchanged, but the upper bound is increased to roughly
$\cs^2 = 1.185$, which is a significant increase. As expected, the
weight of the additional shell remains zero as long as $\cs^2$ remains
in the original interval $1/3 \le \cs^2 \le 2/3$. As soon as $\cs^2$
exceeds $2/3$, the weight of the additional shell starts to increase,
while at the same time the weight of the shell $c_i^2 = 1$ drops to
zero and remains at that value. Therefore we have essentially joined
two minimal models. Indeed, running the main script for the set
$c_i^2 = 2, 4, 5$ results in a unique solution and a range of
validity $2/3 \le \cs^2 \le 32/27$.

\subsection{Three--dimensional models}

For $M = 4$, i.~e. $R = 3$, we were unable to find a suitable velocity
set that would comprise only two non--trivial shells. A
straightforward and simple choice for three shells would be $c_i^2 =
1, 2, 3$, corresponding to typical vectors $(1, 0 , 0)$, $(1, 1, 0)$,
and $(1, 1, 1)$. In this case the matrix turns out to be
rank--deficient, and there is no solution. Trying the three shells
$c_i^2 = 1, 2, 4$ (last value corresponding to a typical vector $(2,
0, 0)$) gives rise to a $25$--speed model (6 / 12 / 6 vectors in the
non--trivial shells) with unique solution
\begin{eqnarray}
  w_{(000)} & = & 1 - \frac{15}{4} \cs^2 + \frac{21}{4} \cs^4 \\
  w_{(100)} & = & \frac{2}{3} \cs^2 - \frac{3}{2} \cs^4 \\
  w_{(110)} & = & \frac{1}{4} \cs^4 \\
  w_{(200)} & = & - \frac{1}{24} \cs^2 + \frac{1}{8} \cs^4 .
\end{eqnarray}
This is valid in the interval $1/3 \le \cs^2 \le 4/9$. At $\cs^2 =
1/3$, the shell $c_i^2 = 4$ may be discarded, such that we recover the
well--known D3Q19 model~\cite{qian_lattice_1992} with $w_{(000)} =
1/3$, $w_{(100)} = 1/18$ and $w_{(110)} = 1/36$.  At $\cs^2 = 4/9$ the
shell $c_i^2 = 1$ can be discarded, giving rise to another 19--speed
model with $w_{(000)} = 10/27$, $w_{(110)} = 4/81$ and $w_{(200)} =
1/162$.

Analyzing the set $c_i^2 = 1, 3, 4$ gives rise to a 21--speed model.
The three non--trivial shells comprise 6, 8, and 6 vectors, respectively,
with typical vectors $(1, 0, 0)$, $(1, 1, 1)$ and $(2, 0, 0)$. This
model has a unique solution
\begin{eqnarray}
  w_{(000)} & = & 1 - \frac{15}{4} \cs^2 + \frac{17}{4} \cs^4 \\
  w_{(100)} & = & \frac{2}{3} \cs^2 - \cs^4 \\
  w_{(111)} & = & \frac{1}{8} \cs^4 \\
  w_{(200)} & = & - \frac{1}{24} \cs^2 + \frac{1}{8} \cs^4
\end{eqnarray}
and a range of validity $1/3 \le \cs^2 \le 2/3$. At $\cs^2 = 1/3$ we
may discard the shell $c_i^2 = 4$ and recover the standard D3Q15
model~\cite{qian_lattice_1992} with $w_{(000)} = 2/9$, $w_{(100)} =
1/9$ and $w_{(111)} = 1/72$. At $\cs^2 = 2/3$ the shell $c_i^2 = 1$
may be discarded, which gives rise to another 15--speed model with
weights $w_{(000)} = 7/18$, $w_{(111)} = 1/18$ and $w_{(200)} = 1/36$.

We now require that the model satisfies the MBCs up to tensor rank $M
= 6$, i.~e. $R = 6$, such that up to six non--trivial shells are
required. A first attempt with the shells $c_i^2 = 1, 2, 3, 4, 5, 6$
results in a rank--deficient matrix with no solution. Enhancing the
model by the additional shells $c_i^2 = 8, 12, 16$ then yields a
solvable but rank--deficient problem with rank six. We should
therefore be able to again remove up to three shells. We first remove
$c_i^2 = 5, 6$ since these shells have as much as $24$ speeds
each. Indeed the solvability remains. Finally we remove $c_i^2 = 8$,
which contains $12$ speeds, and then obtain a unique solution for a
$47$--speed model comprising $c_i^2 = 1, 2, 3, 4, 12, 16$, i.~e.  six
vectors of type $(1, 0, 0)$, twelve of type $(1, 1, 0)$, eight of type
$(1, 1, 1)$, six of type $(2, 0, 0)$, eight of type $(2, 2, 2)$ and
six of type $(4, 0, 0)$. The solution reads
\begin{eqnarray}
  w_{(000)} & = & 1 - \frac{63}{16} \cs^2 + \frac{357}{64} \cs^4
                - \frac{37}{64} \cs^6  \\
  w_{(100)} & = & \frac{32}{45} \cs^2 - \frac{4}{3} \cs^4
               - \frac{1}{3} \cs^6  \\
  w_{(110)} & = & \frac{1}{2} \cs^6  \\
  w_{(111)} & = & \frac{1}{6} \cs^4 - \frac{3}{8} \cs^6  \\
  w_{(200)} & = & - \frac{1}{18} \cs^2 + \frac{3}{16} \cs^4
               - \frac{1}{12} \cs^6 \\
  w_{(222)} & = & - \frac{1}{384} \cs^4 + \frac{1}{128} \cs^6 \\
  w_{(400)} & = & \frac{1}{1440} \cs^2 - \frac{1}{384} \cs^4
               + \frac{1}{384} \cs^6 ;
\end{eqnarray}
this model has positive weights for $0.3510760 \le \cs^2 \le 4/9$. The
former value is irrational and results from $w_{(200)} = 0$; the exact
number is $\cs^2 = 9/8 - \sqrt{115/192}$. At this $\cs^2$ value we may
discard the shell $c_i^2 = 4$, such that we obtain a reduced 41--speed
model with $w_{(000)} = 0.2801500$, $w_{(100)} = 0.07089101$,
$w_{(110)} = 0.02163583$, $w_{(111)} = 4.315525 \times 10^{-3}$,
$w_{(222)} = 1.708487 \times 10^{-5}$ and $w_{(400)} = 3.551447 \times
10^{-5}$. Conversely, for $\cs^2 = 4/9$ we may discard the shell
$c_i^2 = 3$, such that we obtain a 39--speed model with weights
$w_{(000)} = 0.3010974$, $w_{(100)} = 0.02341107$, $w_{(110)} =
0.04389575$, $w_{(200)} = 5.029721 \times 10^{-3}$, $w_{(222)} =
1.714678 \times 10^{-4}$, and $w_{(400)} = 2.286237 \times 10^{-5}$.

It is worth noting that the thus--derived 41--velocity model is
different from the 41--speed model discussed by Chikatamarla and
Karlin~\cite{chikatamarla_lattices_2009}. The latter comprises the
five non--trivial shells $c_i^2 = 1, 2, 3, 9, 27$, where in the case
of $c_i^2 = 9$ only the six vectors of type $(3, 0, 0)$ are taken into
account, while the $c_i^2 = 27$ shell contains only the eight
vectors of type $(3,3,3)$. To analyze this case, we need to add one
more shell in order to allow for a varying $\cs^2$ value, for which we
take $c_i^2 = 16$. Indeed we then find that the model has a unique
solution and a fairly narrow range of validity of
$0.3500280 \le \cs^2 \le 0.3675445$.  At the upper limit the weight of
$c_i^2 = 16$ vanishes, and thus we recover the model of
Ref.~\cite{chikatamarla_lattices_2009}. Here we find
$w_{(000)} = 0.2759976, w_{(100)} = 0.06508547, w_{(110)} =
0.02482560, w_{(111)} = 4.256684 \times 10^{-3}, w_{(300)} = 2.512627
\times 10^{-4}, w_{(333)} = 2.674506 \times 10^{-6}$. Comparison with
Ref.~\cite{chikatamarla_lattices_2009} shows that these parameters are
identical to the numbers given there.

\begin{table}
  \begin{tabular}{| c | c | c | c |}
    \hline
    shell & typical & weight at
    & weight at \\
    size  & vector  & $\cs^2 = 6.979533 \times 10^{-1}$
    & $\cs^2 = 9.470745 \times 10^{-1}$ \\
    \hline
    1  & (0, 0, 0) & $1.543187 \times 10^{-1}$ & $2.350425 \times 10^{-2}$ \\
    6  & (0, 0, 1) & $2.651360 \times 10^{-2}$ & $7.092721 \times 10^{-2}$ \\
    12 & (0, 1, 1) & $4.083040 \times 10^{-2}$ & $1.015888 \times 10^{-4}$ \\
    8  & (1, 1, 1) & $5.220616 \times 10^{-3}$ & $3.488597 \times 10^{-2}$ \\
    6  & (0, 0, 2) & $1.201068 \times 10^{-2}$ & $2.144855 \times 10^{-2}$ \\
    24 & (1, 1, 2) & $2.763355 \times 10^{-3}$ & $2.987112 \times 10^{-3}$ \\
    12 & (0, 2, 2) & $9.685223 \times 10^{-4}$ & $4.073125 \times 10^{-3}$ \\
    6  & (0, 0, 3) & $2.645967 \times 10^{-4}$ & $0                     $ \\
    24 & (1, 1, 3) & $1.362802 \times 10^{-4}$ & $8.608570 \times 10^{-4}$ \\
    6  & (0, 0, 4) & $0                     $ & $9.526366 \times 10^{-5}$ \\
    8  & (3, 3, 3) & $6.029897 \times 10^{-7}$ & $1.674948 \times 10^{-5}$ \\
    \hline
  \end{tabular}
  \caption{Properties of a $113$--speed model in three dimensions
    that is isotropic up to tensor rank $8$.}
  \label{tab:d3m8}
\end{table}
\begin{table}
  \begin{tabular}{| c | c | c | c |}
    \hline
    shell & typical & weight at          & weight at \\
    size  & vector  & $\cs^2 = 1.033691$ & $\cs^2 = 1.206545$ \\
    \hline
    1  & (0, 0, 0) & $1.125792 \times 10^{-1}$ & $5.101845 \times 10^{-2}$ \\
    6  & (0, 0, 1) & $1.444892 \times 10^{-2}$ & $3.953745 \times 10^{-2}$ \\
    12 & (0, 1, 1) & $2.781069 \times 10^{-2}$ & $4.937669 \times 10^{-3}$ \\
    8  & (1, 1, 1) & $1.970138 \times 10^{-2}$ & $3.536908 \times 10^{-2}$ \\
    6  & (0, 0, 2) & $2.251462 \times 10^{-2}$ & $2.485832 \times 10^{-2}$ \\
    24 & (1, 1, 2) & $3.624508 \times 10^{-3}$ & $3.216647 \times 10^{-3}$ \\
    12 & (0, 2, 2) & $4.387148 \times 10^{-3}$ & $7.022298 \times 10^{-3}$ \\
    6  & (0, 0, 3) & $6.910281 \times 10^{-4}$ & $1.578096 \times 10^{-3}$ \\
    24 & (1, 1, 3) & $1.038248 \times 10^{-3}$ & $1.597874 \times 10^{-3}$ \\
    8  & (2, 2, 2) & $4.381319 \times 10^{-4}$ & $5.451840 \times 10^{-4}$ \\
    24 & (0, 1, 4) & $3.513518 \times 10^{-5}$ & $0                     $ \\
    24 & (2, 2, 3) & $4.350915 \times 10^{-5}$ & $1.453046 \times 10^{-4}$ \\
    24 & (1, 1, 4) & $0                     $ & $9.956211 \times 10^{-5}$ \\
    12 & (0, 3, 3) & $1.885761 \times 10^{-6}$ & $3.047305 \times 10^{-5}$ \\
    6  & (0, 0, 5) & $2.394034 \times 10^{-6}$ & $1.300108 \times 10^{-5}$ \\
    24 & (0, 3, 4) & $7.194413 \times 10^{-6}$ & $1.815117 \times 10^{-5}$ \\
    \hline
  \end{tabular}
  \caption{Properties of a $221$--speed model in three dimensions
    that is isotropic up to tensor rank $10$.}
  \label{tab:d3m10}
\end{table}

With some trial and error (along similar lines as described in more
detail for the two--dimensional case), we were also able to find
minimal models for $M = 8$ ($R = 11$) and $M =10$ ($R = 18$). For
eighth--order isotropy, a model of ten non--trivial shells turns out
to be sufficient: $c_i^2 = 1, 2, 3, 4, 6, 8, 9, 11, 16, 27$, where for
$c_i^2 = 9$ we take the subshell of type $(3, 0, 0)$ and for
$c_i^2 = 27$ the subshell of type $(3, 3, 3)$ (all other shells are
irreducible). We thus have a $113$--speed model which is valid in the
interval $0.6979533 \le \cs^2 \le 0.9470745$ and which reduces to a
$107$--speed model at both the upper and the lower end of the interval
of validity.

Similarly, we also found a possible minimal model with tenth--order
isotropy. This is facilitated by the set
$c_i^2 = 1, 2, 3, 4, 6, 8, 9, 11, 12, 17, 18, 25$, where the shell
$c_i^2 = 9$ is restricted to vectors of type $(3, 0, 0)$, while for
all other modulus shells we take all subshells. This set comprises
$221$ velocities in total and the model is valid in the interval
$1.033691 \le \cs^2 \le 1.206545$. The reduced models at the lower and
upper limit of validity are obtained by elimination of shells which
both contain $24$ velocities. The reduced models are therefore both
$197$--speed models.

For these two final models we do not present the expansions of the
weights in powers of $\cs^2$, for similar reasons as for the case $d =
2$, $M = 10$. Other model properties are summarized in
Tabs.~\ref{tab:d3m8} and \ref{tab:d3m10}.

The three--dimensional models that were investigated in a recent paper
by Shan~\cite{shan_pap05_2016} (going up to tensor order $M = 8$)
could all be verified, see Appendix~\ref{sec:ComparisonShan}.

\section{Summary}
\label{sec:summary}

The present study has shown that it is possible to formulate the
problem of constructing weight coefficients in an LB model as one of
numerical linear algebra. Crucial for this to work was (i) the notion
of $\cs^2$ as a free parameter; (ii) a detailed understanding of the
symmetry restrictions on the dimensionality of the underlying tensor
spaces; (iii) a mapping of the tensor equations to scalar equations by
contraction with tensors of the form $n_\alpha n_\beta \ldots
n_\gamma$ constructed from random unit vectors; and (iv) analysis of
the linear--algebra problem in terms of the singular--value
decomposition. Putting these observations into software, it is
possible to write a program that (i) checks for the validity of a
given set of shells, and (ii) calculates the corresponding weights. We
found it encouraging to see with what ease the automatic script does
all the algebra to derive standard LB models and even new ones --- to
the best of our knowledge, so far no LB model has been discussed in
the literature that is isotropic up to tensor rank ten. The successful
examples of Sec.~\ref{sec:examples} show clearly that this is a fairly
useful tool for the LB community.

\begin{acknowledgments}
  Stimulating discussions with J. Zelko, P. Lehnung, U. Schiller,
  A.~J.~C. Ladd, and N. Tretyakov, are gratefully acknowledged. We
  also thank the latter for a critical reading of the manuscript.  We
  are particularly grateful to Mischa Dombrowski, who contributed to
  the development of the software. Funded by the Deutsche
  Forschungsgemeinschaft (DFG, German Research Foundation) - Project
  number 233630050 - TRR 146.
\end{acknowledgments}

\begin{appendix}
\section{Equilibrium populations at non--zero flow velocity}
\label{sec:equilpopul}

\subsection{Polynomials in the flow velocity}

We recall that the original problem of constructing an LB model is not
the fulfillment of MBCs in the absence of flow, as specified in
Eq.~\ref{eq:MBC}, but rather the more general problem of finding
equilibrium populations $n_i^{eq}$ that satisfy an analogous relation
based upon a Maxwell--Boltzmann distribution centered around the local
flow velocity $\vec u$:
\be
  \label{eq:MoreGeneralMBC}
  \rho^{-1} \sum_i n_i^{eq} c_{i \alpha} c_{i \beta} \ldots c_{i \gamma} =
  \int d^d \vec v f(\vec v - \vec u) v_\alpha v_\beta \ldots
  v_\gamma ,
\ee
and we require that this holds for all tensors up to a certain rank
$K$. For example, the populations according to
Eq.~\ref{eq:equilibrium} satisfy Eq.~\ref{eq:MoreGeneralMBC} up to
tensor rank $K = 2$
(cf. Eqs.~\ref{eq:equiltensor0}--\ref{eq:equiltensor2}). However, it
turns out that this problem can be solved fairly easily as soon as the
set of velocities $\vec c_i$, along with the corresponding set of
weights $w_i$, has been found. The MBC problem according to
Eq.~\ref{eq:MBC} must be solved up to tensor rank $M = 2 K$, and then
a straightforward solution of Eq.~\ref{eq:MoreGeneralMBC} is found in
terms of a tensorial polynomial of order $K$ in $\vec u$, where the
expansion coefficients are essentially the tensor Hermite polynomials
in $\vec c_i$, which were introduced into LB theory by He and
Luo~\cite{he_theory_1997}. How this is done will be detailed below. It
thus turns out that the most difficult aspect of the problem is the
identification of a proper set of velocities and the determination of
the weights (as should have become quite clear from the main text).

To simplify the problem of Eq.~\ref{eq:MoreGeneralMBC} we first
introduce suitably scaled variables: $\nu_i = n_i^{eq} / (w_i \rho)$,
$\vec d_i = \vec c_i / \cs$, $\vec \xi = \vec v / \cs$, $\vec \eta =
\vec u / \cs$, as well as a normalized Maxwell--Boltzmann distribution
\be
  \phi (\vec \xi) = (2 \pi)^{-d/2}
  \exp \left( - \frac{\vec \xi^2}{2} \right) .
\ee
In terms of these variables, Eq.~\ref{eq:MoreGeneralMBC} is written as
\be
  \label{eq:MoreGeneralMBCScaled}
  \sum_i w_i \nu_i d_{i \alpha} d_{i \beta} \ldots d_{i \gamma} =
  \int d^d \vec \xi \, \phi(\vec \xi - \vec \eta) \xi_\alpha
  \xi_\beta \ldots \xi_\gamma .
\ee

At this point, it is useful to introduce tensor Hermite
polynomials~\cite{shan_kinetic_2006, grad_note_1949} via their
definition
\be
  H^{(n)}_{\alpha \beta \ldots \gamma} (\vec \xi) = (-1)^n
  \phi(\vec \xi)^{-1} \partial_\alpha \partial_\beta \ldots
  \partial_\gamma \phi(\vec \xi) ,
\ee
where $\partial_\alpha$ denotes a derivative in velocity space,
$\partial_\alpha = \partial / \partial \xi_\alpha$. It should be noted
that $n$ denotes both the rank of the tensor as well as the degree of
the polynomial in $\vec \xi$. It can be shown~\cite{grad_note_1949}
that the polynomials are mutually orthogonal with respect to the
weight function $\phi(\vec \xi)$. The definition implies that the
Taylor expansion of $\phi(\vec \xi - \vec \eta)$ with respect to $\vec
\eta$ reads
\be
  \label{eq:HermiteTaylorExpansion}
  \phi(\vec \xi - \vec \eta) = \sum_{m = 0}^\infty \frac{1}{m!}
  \phi(\vec \xi)
  H^{(m)}_{\mu \sigma \ldots \tau} (\vec \xi)
  \eta_\mu \eta_\sigma \ldots \eta_\tau .
\ee
Now, instead of requiring the identity of tensor moments up to rank
$K$ (Eq.~\ref{eq:MoreGeneralMBCScaled}), we may equivalently require
the identity of the corresponding expressions, where the products
$d_{i \alpha} d_{i \beta} \ldots d_{i \gamma}$ and $\xi_\alpha
\xi_\beta \ldots \xi_\gamma$ are replaced by the corresponding Hermite
polynomials up to order $K$:
\be
  \sum_i w_i \nu_i H^{(n)}_{\alpha \beta \ldots \gamma} (\vec d_i)
  =
  \int d^d \vec \xi \, \phi(\vec \xi - \vec \eta)
  H^{(n)}_{\alpha \beta \ldots \gamma} (\vec \xi) .
\ee
We now insert the Taylor expansion, Eq.~\ref{eq:HermiteTaylorExpansion}.
Making use of orthogonality, one sees that only the term $m = n$
survives:
\begin{eqnarray}
  \label{eq:HermitePolynomialIdentity}
  &&
  \sum_i w_i \nu_i H^{(n)}_{\alpha \beta \ldots \gamma} (\vec d_i)
  \\
  \nonumber
  & = &
  \frac{1}{n!} \int d^d \vec \xi \, \phi(\vec \xi)
  H^{(n)}_{\alpha \beta \ldots \gamma} (\vec \xi)
  H^{(n)}_{\mu \sigma \ldots \tau} (\vec \xi)
  \eta_\mu \eta_\sigma \ldots \eta_\tau .
\end{eqnarray}
For our purposes, it is not necessary to evaluate the rhs
further. Rather we note that Eq.~\ref{eq:HermitePolynomialIdentity}
needs to be satisfied for all $n$ with $0 \le n \le K$, and we now
wish to show that the polynomial ansatz
\be
  \label{eq:PolynomialAnsatz}
  \nu_i = \sum_{m = 0}^K \frac{1}{m!}
  H^{(m)}_{\mu \sigma \ldots \tau} (\vec d_i)
  \eta_\mu \eta_\sigma \ldots \eta_\tau ,
\ee
which is, in essence, a polynomial in the flow velocity $\vec u$, does
indeed solve the problem. Inserting the ansatz into the lhs of
Eq.~\ref{eq:HermitePolynomialIdentity}, one sees that there
polynomials in $\vec d_i$ occur, whose order does not exceed $2 K$.
However, the coefficients $w_i$ have already been adjusted such that
the $\vec u = 0$ MBCs are satisfied up to order $2 K$. It is therefore
justified to replace $\sum_i w_i \ldots$ on the lhs with $\int d^d
\vec \xi \phi(\vec \xi) \ldots$, where simultaneously $\vec d_i$ is
being replaced by $\vec \xi$. Again, orthogonality tells us that only
the term $m = n$ survives. After these operations, it becomes obvious
that rhs and lhs are identical, which completes the proof.

\subsection{The entropic approach}

An alternative approach is to find the equilibrium populations by
maximizing a suitably constructed entropy. This has been popularized
by the so--called ``entropic'' LB method~\cite{karlin_perfect_1999,
  boghosian_galilean-invariant_2003}. The entropy can be derived by
elementary statistical considerations, as outlined in
Ref.~\cite{dunweg_statistical_2007}. Here one assumes a lattice gas
with many particles on each lattice site, such that the notion of a
single--site entropy makes sense. Each particle has a mass $m$, and
we define $\mu$ as the associated mass density, $\mu = m / a^d$, where
$a$ is the lattice spacing. The model then yields for the entropy
\be
  \label{eq:DefineEntropy}
  S = - \sum_i \frac{\rho w_i}{\mu} \left(
    \frac{n_i}{\rho w_i} \ln \frac{n_i}{\rho w_i}
    + 1 - \frac{n_i}{\rho w_i} \right) .
\ee
Defining a scaled entropy as $\tilde{S} = \mu S / \rho$, this can
be written in terms of the reduced variables of the previous
subsection:
\be
  \label{eq:ScaledEntropy}
  \tilde{S} = - \sum_i w_i
  \left( \nu_i \ln \nu_i + 1 - \nu_i \right) .
\ee
The equilibrium populations are then found by maximizing $S$
under the constraints of given mass and momentum,
\begin{eqnarray}
  \sum_i n_i & = & \rho , \\
  \sum_i n_i \vec c_i & = & \rho \vec u ,
\end{eqnarray}
or
\begin{eqnarray}
  \label{eq:MaxEntConstraint1}
  \sum_i w_i \nu_i & = & 1 , \\
  \label{eq:MaxEntConstraint2}
  \sum_i w_i \nu_i \vec d_i & = & \vec \eta .
\end{eqnarray}
Introducing Lagrange multipliers $\lambda_\rho$ and $\vec \lambda_{\vec u}$,
we consider
\be
  \tilde{S}^\prime = \tilde{S} - \lambda_\rho \sum_i w_i \nu_i
  - \vec \lambda_{\vec u} \cdot \sum_i w_i \nu_i \vec d_i .
\ee
The solution of the maximum--entropy problem is then
\be
  \nu_i = \exp \left( - \lambda_\rho
  - \vec \lambda_{\vec u} \cdot \vec d_i \right) ,
\ee
where the Lagrange multipliers must be determined via the constraint
equations, Eqs.~\ref{eq:MaxEntConstraint1} and
\ref{eq:MaxEntConstraint2}:
\begin{eqnarray}
  \label{eq:MaxEntConstraint1New}
  \exp\left(- \lambda_\rho \right)
  \sum_i w_i \exp\left( - \vec \lambda_{\vec u} \cdot \vec d_i \right)
  & = & 1 , \\
  \label{eq:MaxEntConstraint2New}
  \exp\left(- \lambda_\rho \right)
  \sum_i w_i \vec d_i
  \exp\left( - \vec \lambda_{\vec u} \cdot \vec d_i \right)
  & = & \vec \eta ,
\end{eqnarray}
or
\begin{eqnarray}
  \label{eq:MaxEntConstraint1Newer}
  \nu_i & = & \frac{
    \exp\left( - \vec \lambda_{\vec u} \cdot \vec d_i \right)}{
    \sum_j w_j
    \exp\left( - \vec \lambda_{\vec u} \cdot \vec d_j \right)} ,
  \\
  \label{eq:MaxEntConstraint2Newer}
  \vec \eta
  & = &
  \frac{\sum_i w_i \vec d_i
    \exp\left( - \vec \lambda_{\vec u} \cdot \vec d_i \right)}{
    \sum_j w_j
    \exp\left( - \vec \lambda_{\vec u} \cdot \vec d_j \right)} ,
\end{eqnarray}
where Eq.~\ref{eq:MaxEntConstraint2Newer} must typically be solved
numerically to determine $\vec \lambda_{\vec u}$, e.~g. by Newton
iteration.

We now wish to show that the solution derived in the previous
subsection, i.~e. a $K$th--order polynomial in the flow velocity $\vec
u$ (see Eq.~\ref{eq:PolynomialAnsatz}), is an approximate solution of
the maximum--entropy problem for small $\vec u$, correct up to error
terms of order $u^{K + 1}$. Equivalently, we may also show that
Eq.~\ref{eq:PolynomialAnsatz}, evaluated for $K = \infty$, is the
\emph{exact} solution of the maximum--entropy problem, and we will
take that latter approach. The proof is complete as soon as it is
clear that the Lagrange multipliers $\lambda_\rho$ and $\vec
\lambda_{\vec u}$ can be adjusted in such a way that
Eqs.~\ref{eq:MaxEntConstraint1New} and \ref{eq:MaxEntConstraint2New}
hold. Since we assume that the MBCs are satisfied up to infinite order
in $\vec u$, we may however replace the terms $\sum_i w_i \ldots$ on
the lhs by the corresponding integrals $\int d^d \vec \xi \phi(\vec
\xi) \ldots$. We therefore obtain
\begin{eqnarray}
  \label{eq:MaxEntConstraint1EvenNewer}
  \int d^d \vec \xi \, \phi(\vec \xi)
  \exp\left( - \vec \lambda_{\vec u} \cdot \vec \xi \right)
  & = &
  \exp\left( \lambda_\rho \right) ,
  \\
  \label{eq:MaxEntConstraint2EvenNewer}
  \int d^d \vec \xi \, \phi(\vec \xi) \, \vec \xi
  \exp\left( - \vec \lambda_{\vec u} \cdot \vec \xi \right)
  & = &
  \vec \eta \exp\left( \lambda_\rho \right) .
\end{eqnarray}
The Gaussian integrals on the lhs are trivial to evaluate; this yields
\begin{eqnarray}
  \label{eq:MaxEntConstraint1YetNewer}
  \exp \left( \frac{1}{2} \vec \lambda_{\vec u}^2 \right)
  & = &
  \exp\left( \lambda_\rho \right) ,
  \\
  \label{eq:MaxEntConstraint2YetNewer}
  - \vec \lambda_{\vec u}
  \exp \left( \frac{1}{2} \vec \lambda_{\vec u}^2 \right)
  & = &
  \vec \eta \exp\left( \lambda_\rho \right) .
\end{eqnarray}
Therefore a solution for the Lagrange multipliers can indeed be found;
it is simply given by $\vec \lambda_{\vec u} = - \vec \eta$ and
$\lambda_\rho = \vec \eta^2 / 2$.

\vspace{1cm}

\section{Comparison with Ref.
  \onlinecite{shan_pap05_2016}}
  \label{sec:ComparisonShan}

We have used the present Python script in order to verify the results
reported in Ref. \onlinecite{shan_pap05_2016}, in which the author
studies the MBCs within the framework of Gauss--Hermite
quadratures. The quadratures are labeled as $E_{d,n}^M$ where $d$ is
the spacial dimension, $n$ is the number of velocities and $M$ is the
highest tensor order satisfied. For the comparison, note that the
parameter $c$ of Ref. \onlinecite{shan_pap05_2016} must be identified
with $1/\cs$ in the notation of the present paper. Furthermore, it
should be noted that in the present paper ``maximum tensor order''
refers to the largest \emph{non--trivial} (i.~e. even) order, while
the notation of Ref. \onlinecite{shan_pap05_2016} includes the next
tensor order as well (which is trivially satisfied because it is odd).
In other words the notion of, e.~g., ``maximum tensor order 7'' in
Ref.~\onlinecite{shan_pap05_2016} corresponds to ``maximum tensor
order 6'' in the context of the present work.

The ``test'' mode described in Sec.~\ref{sec:test_mode} was written
precisely for such purposes. We used it to check the quoted weights
for one--dimensional (Table 2 of Ref.~\onlinecite{shan_pap05_2016}),
two--dimensional (Tables 3 and 4), and three--dimensional (Tables 5
and 6) models. All numbers given in the paper turned out to be
correct, except for two minor typos, which the script detected by
being unable to verify the weights. The first typo occurs in Table 2,
quadrature $E_{1,7}^9$, where a direct calculation with one additional
(auxiliary) shell $c_i^2 = 16$ shows that the weight $w_4$ for the
shell with $c_i^2 = 9$ should read $812.129$ instead of $8121.29$. The
other typo occurs in Table 4, third model, listed in column six. This
model was checked further by a direct calculation using the velocity
shells $c_i^2 = 1, 2, 4, 5, 9, 13, 18, 16$. Here again $c_i^2 = 16$
serves as an auxiliary shell. We then found that the solution from our
script coincides with the solution from the table, except for the
weight for the typical velocity $(0,2)$ which should read $862.347$
instead of $8623.47$.

We also scrutinized further the quadrature $E_{3, 103}^9$ of
Table 5, last column, by adding the auxiliary shell $(1,1,5)$.
This is a particularly interesting case, since it gives rise
to two disjoint intervals of valid $\cs$ values, and thus to
four distinct models at the boundaries. We have listed these
in Table~\ref{tab:weights_d3m8}; the first coincides with the
results given in Ref.~\onlinecite{shan_pap05_2016}.

\onecolumngrid

\begin{table}
  \begin{tabular}{| c | c | c | c | c | c |}
    \hline
    shell & typical & weight at $\cs^2 = $ &
    weight at $\cs^2 = $ & weight at $\cs^2 = $ &
    weight at $\cs^2 = $ \\
    size  & vector
    & $6.97953322 \times 10^{-1}$
    & $  7.67858981 \times 10^{-1}$
    & $  8.52308171 \times 10^{-1}$
    & $  1.01213280$ \\
    \hline
    1  & (0, 0, 0) & $  3.26333518 \times 10^{-2}$
       & $  3.62888307 \times 10^{-2}$ & $  4.97214340 \times 10^{-2}$
       & $  1.03758046 \times 10^{-1}$ \\
    6  & (0, 0, 1) & $  9.76568336 \times 10^{-2}$
       & $  8.72702806 \times 10^{-2}$ & $  7.28640303 \times 10^{-2}$
       & $  3.78004007 \times 10^{-2}$ \\
    8  & (1, 1, 1) & $  2.80977503 \times 10^{-2}$
       & $  3.12518906 \times 10^{-2}$ & $  3.58424179 \times 10^{-2}$
       & $  4.92746605 \times 10^{-2}$ \\
    6  & (0, 0, 2) & $  1.04525956 \times 10^{-3}$
       & $  4.03636444 \times 10^{-3}$ & $  9.45156051 \times 10^{-3}$
       & $  2.87561664 \times 10^{-2}$ \\
    24 & (0, 1, 2) & $  5.70532902 \times 10^{-3}$
       & $  5.88714307 \times 10^{-3}$ & $  5.23786666 \times 10^{-3}$
       & $0$                           \\
    12 & (0, 2, 2) & $  6.11939270 \times 10^{-4}$
       & $  1.16896856 \times 10^{-3}$ & $  2.18293717 \times 10^{-3}$
       & $  5.49849730 \times 10^{-3}$ \\
    8  & (2, 2, 2) & $  1.55964159 \times 10^{-4}$
       & $  2.85244411 \times 10^{-4}$ & $  4.37068358 \times 10^{-4}$
       & $  6.14662612 \times 10^{-4}$ \\
    6  & (0, 0, 3) & $  2.84443252 \times 10^{-4}$
       & $  3.28336044 \times 10^{-4}$ & $  3.69212708 \times 10^{-4}$
       & $  2.16391171 \times 10^{-4}$ \\
    24 & (1, 1, 3) & $  1.30698376 \times 10^{-4}$
       & $  2.61597860 \times 10^{-4}$ & $  5.00317765 \times 10^{-4}$
       & $  1.26405975 \times 10^{-3}$ \\
    24 & (1, 1, 5) & $0$
       & $  2.83245470 \times 10^{-7}$ & $  9.24300377 \times 10^{-7}$
       & $  4.09498434 \times 10^{-6}$ \\
    8  & (3, 3, 3) & $  1.22319450 \times 10^{-6}$
       & $0$                           & $0$
       & $  8.99234508 \times 10^{-6}$ \\
    \hline
  \end{tabular}
  
  \caption{Properties of a 3-dimensional model that is isotropic up to
    tensor rank 8. There are four distinct speeds of sound at which a
    particular weight vanishes. Depending on which speed of sound is
    chosen this results in either a 103--speed model (model 1 and 4)
    or a 119--speed model (model 2 and 3).}
  
  \label{tab:weights_d3m8}

\end{table}

\twocolumngrid

\section{Accuracy criterion for the ``test'' mode}
\label{sec:AccuracyTestMode}

We are interested in the $i$th component of the residual,
\be
\Delta_i = \sum_j A_{ij} w_j -  b_i
\ee
and wish to check for its vanishing. The matrix $A$ is calculated with
high numerical accuracy, essentially up to machine precision. However,
the weights are input parameters, which are typically given only with
moderate accuracy. We here assume a relative accuracy of $\varepsilon
= 10^{-5}$, such that the (maximum) roundoff error in the
weights is given by
\be
\delta w_j = \varepsilon  w_j .
\ee
Furthermore, the inhomogeneity $b_i$ is subject to a similar lack of
input precision. Recalling that the inhomogeneities are given as
certain powers of $\cs^2$,
\be
b_i = \left( \cs^2 \right)^{m_i/2} ,
\ee
we find that the inaccuracy of $b_i$ is due to the inaccuracy of
$\cs^2$:
\be
\delta b_i = \frac{m_i}{2} \left( \cs^2 \right)^{m_i/2 - 1} \delta \cs^2
\ee
or
\be
\frac{\delta b_i}{b_i} = \frac{m_i}{2} \frac{\delta \cs^2}{\cs^2} .
\ee
Again, the relative accuracy of $\cs^2$ is given by $\varepsilon$; hence
\be
\delta b_i = \varepsilon \frac{m_i}{2} b_i .
\ee
Note that this latter formula is also applicable in the case $b_i =
0$; this situation occurs if the script needs to check the correctness
of a solution that is not given in terms of a single vector $\vec w$
but rather in terms of a whole subspace. From Gaussian error
propagation we then estimate the accuracy of the residual as
\begin{eqnarray}\label{eq:TestThreshold0}
  \nonumber
  \delta \Delta_i
  & = &
  \left[ \sum_j \left( A_{ij} \delta w_j \right)^2 +
    \left( \delta b_i \right)^2 \right]^{1/2}
  \\
  & = &
  \varepsilon 
  \left[ \sum_j \left( A_{ij} w_j \right)^2 +
    \left( \frac{m_i}{2} b_i \right)^2 \right]^{1/2} .
\end{eqnarray}
Whenever $\left\vert \Delta_i \right\vert$ is smaller than this value,
it should be considered to be numerically indistinguishable from zero.

\end{appendix}

\onecolumngrid


\begin{thebibliography}{40}%
\makeatletter
\providecommand \@ifxundefined [1]{%
 \@ifx{#1\undefined}
}%
\providecommand \@ifnum [1]{%
 \ifnum #1\expandafter \@firstoftwo
 \else \expandafter \@secondoftwo
 \fi
}%
\providecommand \@ifx [1]{%
 \ifx #1\expandafter \@firstoftwo
 \else \expandafter \@secondoftwo
 \fi
}%
\providecommand \natexlab [1]{#1}%
\providecommand \enquote  [1]{``#1''}%
\providecommand \bibnamefont  [1]{#1}%
\providecommand \bibfnamefont [1]{#1}%
\providecommand \citenamefont [1]{#1}%
\providecommand \href@noop [0]{\@secondoftwo}%
\providecommand \href [0]{\begingroup \@sanitize@url \@href}%
\providecommand \@href[1]{\@@startlink{#1}\@@href}%
\providecommand \@@href[1]{\endgroup#1\@@endlink}%
\providecommand \@sanitize@url [0]{\catcode `\\12\catcode `\$12\catcode
  `\&12\catcode `\#12\catcode `\^12\catcode `\_12\catcode `\%12\relax}%
\providecommand \@@startlink[1]{}%
\providecommand \@@endlink[0]{}%
\providecommand \url  [0]{\begingroup\@sanitize@url \@url }%
\providecommand \@url [1]{\endgroup\@href {#1}{\urlprefix }}%
\providecommand \urlprefix  [0]{URL }%
\providecommand \Eprint [0]{\href }%
\providecommand \doibase [0]{https://doi.org/}%
\providecommand \selectlanguage [0]{\@gobble}%
\providecommand \bibinfo  [0]{\@secondoftwo}%
\providecommand \bibfield  [0]{\@secondoftwo}%
\providecommand \translation [1]{[#1]}%
\providecommand \BibitemOpen [0]{}%
\providecommand \bibitemStop [0]{}%
\providecommand \bibitemNoStop [0]{.\EOS\space}%
\providecommand \EOS [0]{\spacefactor3000\relax}%
\providecommand \BibitemShut  [1]{\csname bibitem#1\endcsname}%
\let\auto@bib@innerbib\@empty
\bibitem [{\citenamefont {Succi}(2001)}]{succi_lattice_boltzmann_2001}%
  \BibitemOpen
  \bibfield  {author} {\bibinfo {author} {\bibfnamefont {S.}~\bibnamefont
  {Succi}},\ }\href@noop {} {\emph {\bibinfo {title} {The {Lattice} {Boltzmann}
  equation: {For} fluid dynamics and beyond}}}\ (\bibinfo  {publisher}
  {{Oxford} {University} {Press}},\ \bibinfo {address} {Oxford},\ \bibinfo
  {year} {2001})\BibitemShut {NoStop}%
\bibitem [{\citenamefont {Succi}(2018)}]{succi_lattice_2018}%
  \BibitemOpen
  \bibfield  {author} {\bibinfo {author} {\bibfnamefont {S.}~\bibnamefont
  {Succi}},\ }\href@noop {} {\emph {\bibinfo {title} {The {Lattice} {Boltzmann}
  {equation}: {For} {complex} {states} of {flowing} {matter}}}}\ (\bibinfo
  {publisher} {{Oxford} {University} {Press}},\ \bibinfo {address} {Oxford},\
  \bibinfo {year} {2018})\BibitemShut {NoStop}%
\bibitem [{\citenamefont {Kr{\"u}ger}\ \emph {et~al.}(2016)\citenamefont
  {Kr{\"u}ger}, \citenamefont {Kusumaatmaja}, \citenamefont {Kuzmin},
  \citenamefont {Shardt}, \citenamefont {Silva},\ and\ \citenamefont
  {Viggen}}]{kruger_lattice_2016}%
  \BibitemOpen
  \bibfield  {author} {\bibinfo {author} {\bibfnamefont {T.}~\bibnamefont
  {Kr{\"u}ger}}, \bibinfo {author} {\bibfnamefont {H.}~\bibnamefont
  {Kusumaatmaja}}, \bibinfo {author} {\bibfnamefont {A.}~\bibnamefont
  {Kuzmin}}, \bibinfo {author} {\bibfnamefont {O.}~\bibnamefont {Shardt}},
  \bibinfo {author} {\bibfnamefont {G.}~\bibnamefont {Silva}},\ and\ \bibinfo
  {author} {\bibfnamefont {E.~M.}\ \bibnamefont {Viggen}},\ }\href@noop {}
  {\emph {\bibinfo {title} {The {Lattice} {Boltzmann} method: {Principles} and
  practice}}}\ (\bibinfo  {publisher} {Springer},\ \bibinfo {year}
  {2016})\BibitemShut {NoStop}%
\bibitem [{\citenamefont {Benzi}\ \emph {et~al.}(1992)\citenamefont {Benzi},
  \citenamefont {Succi},\ and\ \citenamefont
  {Vergassola}}]{benzi_lattice_1992}%
  \BibitemOpen
  \bibfield  {author} {\bibinfo {author} {\bibfnamefont {R.}~\bibnamefont
  {Benzi}}, \bibinfo {author} {\bibfnamefont {S.}~\bibnamefont {Succi}},\ and\
  \bibinfo {author} {\bibfnamefont {M.}~\bibnamefont {Vergassola}},\
  }\href@noop {} {\bibfield  {journal} {\bibinfo  {journal} {Physics Reports}\
  }\textbf {\bibinfo {volume} {222}},\ \bibinfo {pages} {145} (\bibinfo {year}
  {1992})}\BibitemShut {NoStop}%
\bibitem [{\citenamefont {D{\"u}nweg}\ and\ \citenamefont
  {Ladd}(2009)}]{dunweg_lattice_2009}%
  \BibitemOpen
  \bibfield  {author} {\bibinfo {author} {\bibfnamefont {B.}~\bibnamefont
  {D{\"u}nweg}}\ and\ \bibinfo {author} {\bibfnamefont {A.~J.~C.}\ \bibnamefont
  {Ladd}},\ }in\ \href@noop {} {\emph {\bibinfo {booktitle} {Advanced
  {Computer} {Simulation} {Approaches} for {Soft} {Matter} {Sciences}
  {III}}}},\ \bibinfo {series and number} {\bibinfo {series} {Advances in
  {Polymer} {Science}}\ No.\ \bibinfo {number} {221}},\ \bibinfo {editor}
  {edited by\ \bibinfo {editor} {\bibfnamefont {C.}~\bibnamefont {Holm}}\ and\
  \bibinfo {editor} {\bibfnamefont {K.}~\bibnamefont {Kremer}}}\ (\bibinfo
  {publisher} {Springer},\ \bibinfo {address} {Berlin Heidelberg},\ \bibinfo
  {year} {2009})\ pp.\ \bibinfo {pages} {89--166}\BibitemShut {NoStop}%
\bibitem [{\citenamefont {Higuera}\ \emph {et~al.}(1989)\citenamefont
  {Higuera}, \citenamefont {Succi},\ and\ \citenamefont
  {Benzi}}]{higuera_lattice_1989}%
  \BibitemOpen
  \bibfield  {author} {\bibinfo {author} {\bibfnamefont {F.~J.}\ \bibnamefont
  {Higuera}}, \bibinfo {author} {\bibfnamefont {S.}~\bibnamefont {Succi}},\
  and\ \bibinfo {author} {\bibfnamefont {R.}~\bibnamefont {Benzi}},\
  }\href@noop {} {\bibfield  {journal} {\bibinfo  {journal} {Europhysics
  Letters}\ }\textbf {\bibinfo {volume} {9}},\ \bibinfo {pages} {345} (\bibinfo
  {year} {1989})}\BibitemShut {NoStop}%
\bibitem [{\citenamefont {Qian}\ \emph {et~al.}(1992)\citenamefont {Qian},
  \citenamefont {D'Humieres},\ and\ \citenamefont
  {Lallemand}}]{qian_lattice_1992}%
  \BibitemOpen
  \bibfield  {author} {\bibinfo {author} {\bibfnamefont {Y.~H.}\ \bibnamefont
  {Qian}}, \bibinfo {author} {\bibfnamefont {D.}~\bibnamefont {D'Humieres}},\
  and\ \bibinfo {author} {\bibfnamefont {P.}~\bibnamefont {Lallemand}},\
  }\href@noop {} {\bibfield  {journal} {\bibinfo  {journal} {Europhysics
  Letters}\ }\textbf {\bibinfo {volume} {17}},\ \bibinfo {pages} {479}
  (\bibinfo {year} {1992})}\BibitemShut {NoStop}%
\bibitem [{\citenamefont {Shan}\ \emph {et~al.}(2006)\citenamefont {Shan},
  \citenamefont {Yuan},\ and\ \citenamefont {Chen}}]{shan_kinetic_2006}%
  \BibitemOpen
  \bibfield  {author} {\bibinfo {author} {\bibfnamefont {X.}~\bibnamefont
  {Shan}}, \bibinfo {author} {\bibfnamefont {X.-F.}\ \bibnamefont {Yuan}},\
  and\ \bibinfo {author} {\bibfnamefont {H.}~\bibnamefont {Chen}},\ }\href@noop
  {} {\bibfield  {journal} {\bibinfo  {journal} {Journal of Fluid Mechanics}\
  }\textbf {\bibinfo {volume} {550}},\ \bibinfo {pages} {413} (\bibinfo {year}
  {2006})}\BibitemShut {NoStop}%
\bibitem [{\citenamefont {Philippi}\ \emph {et~al.}(2006)\citenamefont
  {Philippi}, \citenamefont {Hegele}, \citenamefont {dos Santos},\ and\
  \citenamefont {Surmas}}]{philippi_continuous_2006}%
  \BibitemOpen
  \bibfield  {author} {\bibinfo {author} {\bibfnamefont {P.~C.}\ \bibnamefont
  {Philippi}}, \bibinfo {author} {\bibfnamefont {L.~A.}\ \bibnamefont
  {Hegele}}, \bibinfo {author} {\bibfnamefont {L.~O.~E.}\ \bibnamefont {dos
  Santos}},\ and\ \bibinfo {author} {\bibfnamefont {R.}~\bibnamefont
  {Surmas}},\ }\href@noop {} {\bibfield  {journal} {\bibinfo  {journal}
  {Physical Review E}\ }\textbf {\bibinfo {volume} {73}},\ \bibinfo {pages}
  {056702} (\bibinfo {year} {2006})}\BibitemShut {NoStop}%
\bibitem [{\citenamefont {Chen}\ \emph {et~al.}(2008)\citenamefont {Chen},
  \citenamefont {Goldhirsch},\ and\ \citenamefont
  {Orszag}}]{chen_discrete_2008}%
  \BibitemOpen
  \bibfield  {author} {\bibinfo {author} {\bibfnamefont {H.}~\bibnamefont
  {Chen}}, \bibinfo {author} {\bibfnamefont {I.}~\bibnamefont {Goldhirsch}},\
  and\ \bibinfo {author} {\bibfnamefont {S.~A.}\ \bibnamefont {Orszag}},\
  }\href@noop {} {\bibfield  {journal} {\bibinfo  {journal} {Journal of
  Scientific Computing}\ }\textbf {\bibinfo {volume} {34}},\ \bibinfo {pages}
  {87} (\bibinfo {year} {2008})}\BibitemShut {NoStop}%
\bibitem [{\citenamefont {Chikatamarla}\ and\ \citenamefont
  {Karlin}(2009)}]{chikatamarla_lattices_2009}%
  \BibitemOpen
  \bibfield  {author} {\bibinfo {author} {\bibfnamefont {S.~S.}\ \bibnamefont
  {Chikatamarla}}\ and\ \bibinfo {author} {\bibfnamefont {I.~V.}\ \bibnamefont
  {Karlin}},\ }\href@noop {} {\bibfield  {journal} {\bibinfo  {journal}
  {Physical Review E}\ }\textbf {\bibinfo {volume} {79}},\ \bibinfo {pages}
    {046701} (\bibinfo {year} {2009})}\BibitemShut {NoStop}%
\bibitem [{\citenamefont {Karlin}\ and\ \citenamefont
  {Asinari}(2010)}]{karlin_pap07_2010}%
  \BibitemOpen
  \bibfield  {author} {\bibinfo {author} {\bibfnamefont {I.}~\bibnamefont
  {Karlin}}\ and\ \bibinfo {author} {\bibfnamefont {P.}~\bibnamefont
  {Asinari}},\ }\href {https://doi.org/10.1016/j.physa.2009.12.032} {\bibfield
  {journal} {\bibinfo  {journal} {Physica A: Statistical Mechanics and its
  Applications}\ }\textbf {\bibinfo {volume} {389}},\ \bibinfo {pages} {1530}
  (\bibinfo {year} {2010})}\BibitemShut {NoStop}%
\bibitem [{\citenamefont {Shan}(2010)}]{shan_pap04_2010}%
  \BibitemOpen
  \bibfield  {author} {\bibinfo {author} {\bibfnamefont {X.}~\bibnamefont
  {Shan}},\ }\href@noop {} {\bibfield  {journal} {\bibinfo  {journal} {Physical
  Review E}\ }\textbf {\bibinfo {volume} {81}},\ \bibinfo {pages} {036702}
    (\bibinfo {year} {2010})}\BibitemShut {NoStop}%
\bibitem [{\citenamefont {Shan}(2016)}]{shan_pap05_2016}%
  \BibitemOpen
  \bibfield  {author} {\bibinfo {author} {\bibfnamefont {X.}~\bibnamefont
  {Shan}},\ }\href@noop {} {\bibfield  {journal} {\bibinfo  {journal} {Journal
  of Computational Science}\ }\bibinfo {series} {Discrete {Simulation} of
  {Fluid} {Dynamics} 2015},\ \textbf {\bibinfo {volume} {17}},\ \bibinfo
    {pages} {475} (\bibinfo {year} {2016})}\BibitemShut {NoStop}%
\bibitem [{\citenamefont {Shan}(2019)}]{shan_central-moment-based_2019}%
  \BibitemOpen
  \bibfield  {author} {\bibinfo {author} {\bibfnamefont {X.}~\bibnamefont
  {Shan}},\ }\href {https://doi.org/10.1103/PhysRevE.100.043308} {\bibfield
  {journal} {\bibinfo  {journal} {Physical Review E}\ }\textbf {\bibinfo
  {volume} {100}},\ \bibinfo {pages} {043308} (\bibinfo {year}
  {2019})}\BibitemShut {NoStop}%
\bibitem [{\citenamefont {White}\ and\ \citenamefont
  {Chong}(2011)}]{white_pap02_2011}%
  \BibitemOpen
  \bibfield  {author} {\bibinfo {author} {\bibfnamefont {A.~T.}\ \bibnamefont
  {White}}\ and\ \bibinfo {author} {\bibfnamefont {C.~K.}\ \bibnamefont
  {Chong}},\ }\href {https://doi.org/10.1016/j.jcp.2011.04.031} {\bibfield
  {journal} {\bibinfo  {journal} {Journal of Computational Physics}\ }\textbf
  {\bibinfo {volume} {230}},\ \bibinfo {pages} {6367} (\bibinfo {year}
  {2011})}\BibitemShut {NoStop}%
\bibitem [{\citenamefont {Silva}\ and\ \citenamefont
  {Semiao}(2014)}]{silva_pap03_2014}%
  \BibitemOpen
  \bibfield  {author} {\bibinfo {author} {\bibfnamefont {G.}~\bibnamefont
  {Silva}}\ and\ \bibinfo {author} {\bibfnamefont {V.}~\bibnamefont {Semiao}},\
  }\href {https://doi.org/10.1016/j.jcp.2014.03.027} {\bibfield  {journal}
  {\bibinfo  {journal} {Journal of Computational Physics}\ }\textbf {\bibinfo
  {volume} {269}},\ \bibinfo {pages} {259} (\bibinfo {year}
  {2014})}\BibitemShut {NoStop}%
\bibitem [{\citenamefont {Wagner}(2006)}]{wagner_thermodynamic_2006}%
  \BibitemOpen
  \bibfield  {author} {\bibinfo {author} {\bibfnamefont {A.~J.}\ \bibnamefont
  {Wagner}},\ }\href@noop {} {\bibfield  {journal} {\bibinfo  {journal}
  {Physical Review E}\ }\textbf {\bibinfo {volume} {74}},\ \bibinfo {pages}
  {056703} (\bibinfo {year} {2006})}\BibitemShut {NoStop}%
\bibitem [{pyt(2019{\natexlab{a}})}]{python}%
  \BibitemOpen
  \href@noop {} {\bibinfo {title} {Welcome to {Python}.org}},\ \bibinfo
  {howpublished} {\url{https://www.python.org/}} (\bibinfo {year}
  {2019}{\natexlab{a}})\BibitemShut {NoStop}%
\bibitem [{lbw(2019)}]{lbweightscode}%
  \BibitemOpen
  \href@noop {} {\bibinfo {title} {{Lattice}-{Boltzmann}-weights}},\ \bibinfo
  {howpublished} {\url{https://github.com/BDuenweg/Lattice-Boltzmann-weights}}
  (\bibinfo {year} {2019})\BibitemShut {NoStop}%
\bibitem [{num(2019{\natexlab{a}})}]{numpy}%
  \BibitemOpen
  \href@noop {} {\bibinfo {title} {Numpy}},\ \bibinfo {howpublished}
  {\url{http://www.numpy.org/}} (\bibinfo {year}
  {2019}{\natexlab{a}})\BibitemShut {NoStop}%
\bibitem [{num(2019{\natexlab{b}})}]{numpy_linear_algebra}%
  \BibitemOpen
  \href@noop {} {\bibinfo {title} {Linear algebra (numpy.linalg)}},\ \bibinfo
  {howpublished}
  {\url{https://docs.scipy.org/doc/numpy/reference/routines.linalg.html}}
  (\bibinfo {year} {2019}{\natexlab{b}})\BibitemShut {NoStop}%
\bibitem [{wik(2019)}]{wikipedia_partition_2018}%
  \BibitemOpen
  \href@noop {} {\bibinfo {title} {Partition (number theory)}},\ \bibinfo
  {howpublished}
  {\url{https://en.wikipedia.org/wiki/Partition_(number_theory)}} (\bibinfo
  {year} {2019})\BibitemShut {NoStop}%
\bibitem [{gau(2019)}]{gaussian_2017}%
  \BibitemOpen
  \href@noop {} {\bibinfo {title} {Gaussian integral}},\ \bibinfo
  {howpublished} {\url{https://en.wikipedia.org/wiki/Gaussian_integral}}
  (\bibinfo {year} {2019})\BibitemShut {NoStop}%
\bibitem [{\citenamefont {Press}\ \emph {et~al.}(2007)\citenamefont {Press},
  \citenamefont {Teukolsky}, \citenamefont {Vetterling},\ and\ \citenamefont
  {Flannery}}]{press_numerical_2007}%
  \BibitemOpen
  \bibfield  {author} {\bibinfo {author} {\bibfnamefont {W.~H.}\ \bibnamefont
  {Press}}, \bibinfo {author} {\bibfnamefont {S.~A.}\ \bibnamefont
  {Teukolsky}}, \bibinfo {author} {\bibfnamefont {W.~T.}\ \bibnamefont
  {Vetterling}},\ and\ \bibinfo {author} {\bibfnamefont {B.~P.}\ \bibnamefont
  {Flannery}},\ }\href@noop {} {\emph {\bibinfo {title} {Numerical {Recipes}
  3rd Edition: {The} art of scientific computing}}},\ \bibinfo {edition} {3rd}\
  ed.\ (\bibinfo  {publisher} {Cambridge University Press},\ \bibinfo {address}
  {Cambridge, UK ; New York},\ \bibinfo {year} {2007})\BibitemShut {NoStop}%
\bibitem [{sin(2019)}]{singular_2017}%
  \BibitemOpen
  \href@noop {} {\bibinfo {title} {Singular value decomposition}},\ \bibinfo
  {howpublished}
  {\url{https://en.wikipedia.org/wiki/Singular_value_decomposition}} (\bibinfo
  {year} {2019})\BibitemShut {NoStop}%
\bibitem [{num(2019{\natexlab{c}})}]{numpy_svd}%
  \BibitemOpen
  \href@noop {} {\bibinfo {title} {numpy.linalg.svd}},\ \bibinfo {howpublished}
  {\url{https://docs.scipy.org/doc/numpy/reference/generated/numpy.linalg.svd.html}}
  (\bibinfo {year} {2019}{\natexlab{c}})\BibitemShut {NoStop}%
\bibitem [{num(2019{\natexlab{d}})}]{numpy_norm}%
  \BibitemOpen
  \href@noop {} {\bibinfo {title} {numpy.linalg.norm}},\ \bibinfo
  {howpublished}
  {\url{https://docs.scipy.org/doc/numpy/reference/generated/numpy.linalg.norm.html}}
  (\bibinfo {year} {2019}{\natexlab{d}})\BibitemShut {NoStop}%
\bibitem [{num(2019{\natexlab{e}})}]{numpy_roots}%
  \BibitemOpen
  \href@noop {} {\bibinfo {title} {numpy.roots}},\ \bibinfo {howpublished}
  {\url{https://docs.scipy.org/doc/numpy/reference/generated/numpy.roots.html}}
  (\bibinfo {year} {2019}{\natexlab{e}})\BibitemShut {NoStop}%
\bibitem [{num(2019{\natexlab{f}})}]{numpy_sort}%
  \BibitemOpen
  \href@noop {} {\bibinfo {title} {numpy.sort}},\ \bibinfo {howpublished}
  {\url{https://docs.scipy.org/doc/numpy/reference/generated/numpy.sort.html}}
  (\bibinfo {year} {2019}{\natexlab{f}})\BibitemShut {NoStop}%
\bibitem [{con(2019)}]{continued_2017}%
  \BibitemOpen
  \href@noop {} {\bibinfo {title} {Continued fraction}},\ \bibinfo
  {howpublished} {\url{https://en.wikipedia.org/wiki/Continued_fraction}}
  (\bibinfo {year} {2019})\BibitemShut {NoStop}%
\bibitem [{pyt(2019{\natexlab{b}})}]{python_fractions}%
  \BibitemOpen
  \href@noop {} {\bibinfo {title} {9.5. fractions - rational numbers}},\
  \bibinfo {howpublished}
  {\url{https://docs.python.org/2/library/fractions.html}} (\bibinfo {year}
  {2019}{\natexlab{b}})\BibitemShut {NoStop}%
\bibitem [{wol(2019)}]{wolframalpha}%
  \BibitemOpen
  \href@noop {} {\bibinfo {title} {Wolfram {Alpha}: {Computational} {Knowledge}
  {Engine}}},\ \bibinfo {howpublished} {\url{https://www.wolframalpha.com/}}
  (\bibinfo {year} {2019})\BibitemShut {NoStop}%
\bibitem [{\citenamefont {Schrijver}(1998)}]{schrijver_theory_1998}%
  \BibitemOpen
  \bibfield  {author} {\bibinfo {author} {\bibfnamefont {A.}~\bibnamefont
  {Schrijver}},\ }\href@noop {} {\emph {\bibinfo {title} {Theory of linear and
  integer programming}}}\ (\bibinfo  {publisher} {John Wiley \& Sons},\
  \bibinfo {year} {1998})\BibitemShut {NoStop}%
\bibitem [{cvx(2019)}]{cvxpy_ref}%
  \BibitemOpen
  \href@noop {} {\bibinfo {title} {Welcome to {CVXPY} 1.0}},\ \bibinfo
  {howpublished} {\url{http://www.cvxpy.org/}} (\bibinfo {year}
  {2019})\BibitemShut {NoStop}%
\bibitem [{\citenamefont {He}\ and\ \citenamefont
  {Luo}(1997)}]{he_theory_1997}%
  \BibitemOpen
  \bibfield  {author} {\bibinfo {author} {\bibfnamefont {X.}~\bibnamefont
  {He}}\ and\ \bibinfo {author} {\bibfnamefont {L.-S.}\ \bibnamefont {Luo}},\
  }\href@noop {} {\bibfield  {journal} {\bibinfo  {journal} {Physical Review
  E}\ }\textbf {\bibinfo {volume} {56}},\ \bibinfo {pages} {6811} (\bibinfo
  {year} {1997})}\BibitemShut {NoStop}%
\bibitem [{\citenamefont {Grad}(1949)}]{grad_note_1949}%
  \BibitemOpen
  \bibfield  {author} {\bibinfo {author} {\bibfnamefont {H.}~\bibnamefont
  {Grad}},\ }\href@noop {} {\bibfield  {journal} {\bibinfo  {journal}
  {Communications on Pure and Applied Mathematics}\ }\textbf {\bibinfo {volume}
  {2}},\ \bibinfo {pages} {325} (\bibinfo {year} {1949})}\BibitemShut {NoStop}%
\bibitem [{\citenamefont {Karlin}\ \emph {et~al.}(1999)\citenamefont {Karlin},
  \citenamefont {Ferrante},\ and\ \citenamefont
  {{\"O}ttinger}}]{karlin_perfect_1999}%
  \BibitemOpen
  \bibfield  {author} {\bibinfo {author} {\bibfnamefont {I.~V.}\ \bibnamefont
  {Karlin}}, \bibinfo {author} {\bibfnamefont {A.}~\bibnamefont {Ferrante}},\
  and\ \bibinfo {author} {\bibfnamefont {H.~C.}\ \bibnamefont {{\"O}ttinger}},\
  }\href@noop {} {\bibfield  {journal} {\bibinfo  {journal} {Europhysics
  Letters}\ }\textbf {\bibinfo {volume} {47}},\ \bibinfo {pages} {182}
  (\bibinfo {year} {1999})}\BibitemShut {NoStop}%
\bibitem [{\citenamefont {Boghosian}\ \emph {et~al.}(2003)\citenamefont
  {Boghosian}, \citenamefont {Love}, \citenamefont {Coveney}, \citenamefont
  {Karlin}, \citenamefont {Succi},\ and\ \citenamefont
  {Yepez}}]{boghosian_galilean-invariant_2003}%
  \BibitemOpen
  \bibfield  {author} {\bibinfo {author} {\bibfnamefont {B.~M.}\ \bibnamefont
  {Boghosian}}, \bibinfo {author} {\bibfnamefont {P.~J.}\ \bibnamefont {Love}},
  \bibinfo {author} {\bibfnamefont {P.~V.}\ \bibnamefont {Coveney}}, \bibinfo
  {author} {\bibfnamefont {I.~V.}\ \bibnamefont {Karlin}}, \bibinfo {author}
  {\bibfnamefont {S.}~\bibnamefont {Succi}},\ and\ \bibinfo {author}
  {\bibfnamefont {J.}~\bibnamefont {Yepez}},\ }\href@noop {} {\bibfield
  {journal} {\bibinfo  {journal} {Physical Review E}\ }\textbf {\bibinfo
  {volume} {68}},\ \bibinfo {pages} {025103(R)} (\bibinfo {year}
  {2003})}\BibitemShut {NoStop}%
\bibitem [{\citenamefont {D{\"u}nweg}\ \emph {et~al.}(2007)\citenamefont
  {D{\"u}nweg}, \citenamefont {Schiller},\ and\ \citenamefont
  {Ladd}}]{dunweg_statistical_2007}%
  \BibitemOpen
  \bibfield  {author} {\bibinfo {author} {\bibfnamefont {B.}~\bibnamefont
  {D{\"u}nweg}}, \bibinfo {author} {\bibfnamefont {U.~D.}\ \bibnamefont
  {Schiller}},\ and\ \bibinfo {author} {\bibfnamefont {A.~J.~C.}\ \bibnamefont
  {Ladd}},\ }\href@noop {} {\bibfield  {journal} {\bibinfo  {journal} {Physical
  Review E}\ }\textbf {\bibinfo {volume} {76}},\ \bibinfo {pages} {036704}
  (\bibinfo {year} {2007})}\BibitemShut {NoStop}%
\end{thebibliography}
\end{document}